\documentclass[journal]{IEEEtran}
\usepackage{amsmath,amsfonts}
\usepackage{array}
\usepackage{textcomp}
\usepackage{stfloats}
\usepackage{url}
\usepackage{verbatim}
\usepackage{graphicx}
\usepackage{amsmath}
\usepackage{amssymb}
\usepackage{bm}
\usepackage{pifont}
\usepackage{mathtools}
\usepackage{graphicx}
\usepackage{pgfplots}
\pgfplotsset{compat=1.18}
\usepackage{pgfplotstable}
\usepackage{multirow} 
\usepackage{caption}
\usepackage{subcaption}
\usepackage{url}
\usepackage{verbatim}
\usepackage{graphicx}
\usepackage{algorithm}
\usepackage{algpseudocode}      
\usepackage{enumitem}
\usepackage[colorlinks]{hyperref}
\usepackage[dvipsnames]{xcolor}
\hypersetup{
    colorlinks = true,
    linkcolor  = blue,       
    citecolor  = blue,       
    urlcolor   = blue        
}

\newcommand{\CB}{\color{black}}
\def\BibTeX{{\rm B\kern-.05em{\sc i\kern-.025em b}\kern-.08em
    T\kern-.1667em\lower.7ex\hbox{E}\kern-.125emX}}
\usepackage{balance}
\begin{document}
\title{Deep-Unfolded Wideband ISAC Beamforming for DMA Under Frequency-Selective Lorentzian Model}
\author{Abdolrasoul Sakhaei Gharagezlou, {\it Graduate Student Member, IEEE}, Pouya Mobaraki, {\it Graduate Student Member, IEEE},
Mehdi Monemi, {\it Member, IEEE}, Nhan Thanh Nguyen, {\it Senior Member, IEEE},
Mehdi Rasti, {\it Senior Member, IEEE}, Samad Ali, {\it Member, IEEE}, Matti Latva-aho, {\it Fellow, IEEE}
\thanks{Abdolrasoul Sakhaei Gharagezlou, Pouya Mobaraki, Mehdi Monemi, Nhan T. Nguyen, Mehdi Rasti, Samad Ali and Matti Latva-aho are with Centre for Wireless Communications, University of Oulu, Finland (e-mail: \{abdolrasoul.sakhaeigharagezlou; pouya.mobaraki; mehdi.monemi; nhan.nguyen; mehdi.rasti; samad.ali; matti.latva-aho\}@oulu.fi).}}


\markboth{}%
{How to Use the IEEEtran \LaTeX \ Templates}

\maketitle
\begin{abstract}
Integrated sensing and communications (ISAC), empowered by dynamic metasurface antennas (DMAs), has emerged as a promising paradigm for next-generation wireless networks. However, existing DMA-based designs commonly rely on the frequency-flat response model for DMA elements, which is accurate only in narrowband scenarios and can cause significant phase and magnitude mismatches in wideband and ultra-wideband systems. This paper investigates a DMA-based wideband ISAC system under a frequency-selective Lorentzian response model, which accurately captures the frequency-dependent behavior of DMA elements. We aim to jointly balance the aggregate signal-to-interference-plus-noise ratio (SINR) of communication users and the signal-to-noise ratio (SNR) of the radar target. To this end, we first develop an alternating optimization framework based on projected gradient ascent (PGA), deriving closed-form gradients of the objective function with respect to the digital beamforming vectors, resonance frequencies, and damping factors under the frequency-selective Lorentzian DMA model. We then propose an unfolded PGA architecture that preserves the interpretability of model-based optimization while learning key hyperparameters to accelerate convergence. Simulation results show that the frequency-selective Lorentzian model improves performance by approximately 20\% over its frequency-flat approximation. Moreover, deep-unfolded PGA achieves up to 20-fold faster convergence and improves the objective value by up to 7\% compared with PGA-based benchmarks.
\end{abstract}


\section{Introduction}
\IEEEPARstart{N}{ext}-generation wireless systems must support a growing number of stationary and mobile users, provide high data rates, and enable sensing and cognition capabilities \cite{ID1}. Integrated sensing and communication (ISAC) integrates radar sensing and wireless communication capabilities within a unified system infrastructure, offering an efficient solution to meet these growing demands. To further improve performance, technologies such as millimeter-wave and terahertz bands, along with extremely large antenna arrays, are expected to be widely adopted. These technologies provide larger bandwidths, reduce spectrum congestion, enable highly directional beams, and help overcome propagation losses \cite{ID2}. Through the joint use of hardware resources and frequency bands, ISAC systems can improve spectrum efficiency while simultaneously lowering hardware complexity, energy consumption, and deployment costs \cite{ID3}. The development of ISAC has motivated the adaptation of various beamforming architectures, but existing solutions face tradeoffs in beamforming flexibility, hardware cost, power consumption, and scalability \cite{ID4, ID5, ID6, ID7}. This highlights the need for antenna architectures that better balance performance, complexity, and energy efficiency.

Dynamic metasurface antennas (DMAs) are a promising technology for realizing large-scale antenna arrays in a controllable, scalable, and hardware-efficient manner \cite{ID8, ID9}. They use metamaterial radiating elements embedded along microstrip waveguides, where each waveguide is connected to a single radio-frequency (RF) chain \cite{ID10}. This architecture enables compact, planar, and low-cost antenna implementations while naturally supporting hybrid beamforming with far fewer RF chains than fully digital architectures \cite{ID11}. In addition to advances in hardware architectures and ISAC systems, combining these beamforming architectures with wideband and ultra-wideband transmissions provides further benefits, such as enhanced robustness and improved security. These benefits support more reliable communications and greater resilience to interference and jamming \cite{Monemi1, Monemi2}. {\it Consequently, incorporating DMA architectures into wideband ISAC systems facilitates energy- and cost-efficient deployment while enhancing reliability and robustness, thereby supporting the development of high-performance wireless communication networks.}\CB

\subsection{Related Work}
Recent studies have investigated beamforming design for DMA-based architectures in narrowband far-field and near-field communication scenarios \cite{ID11, ID16, ID17, ID19, ID20, ID20.1, SM7}. However, relatively few studies have considered DMA-based ISAC systems, which require the joint optimization of sensing and communication functionalities \cite{SM2, ID13, ID14}. In \cite{SM2}, the authors developed a DMA-based ISAC system for narrowband transmission, designed to maximize radar signal-to-noise ratio (SNR) using sequential rank-one constraint relaxation combined with penalty dual decomposition techniques. The authors in \cite{ID13} considered an in-band full-duplex ISAC system, where the base station (BS), equipped with a DMA-based architecture, communicates with multiple users and detects a sensing target positioned randomly within its coverage area.  In \cite{ID14}, the authors proposed a reconfigurable intelligent surface-assisted near-field ISAC system with symbiotic radio communication, where the BS employs both fully digital and DMA-based antenna architectures to maximize the system’s symbiotic transmission rate. However, these works typically rely on the frequency-flat Lorentzian model, which is only valid for DMA architectures in narrowband scenarios \cite{SM4}. 
 
 In wideband systems, the frequency-flat Lorentzian model for DMA elements restricts element tunability mainly to phase adjustment while neglecting amplitude control. In contrast, the nonlinear {\it frequency-selective} Lorentzian model offers a more accurate characterization of the elements’ frequency-dependent behavior, allowing both phase and magnitude control to be directly integrated into the resource management process. For example, in \cite{SM4}, the authors considered the configurable frequency-selective profile for DMA metamaterial elements in multiple input multiple output (MIMO)-orthogonal frequency-division multiplexing (OFDM) receivers with bit-constrained analog-to-digital converters, resulting in a spectrally flexible hybrid architecture. In \cite{Rasoul}, the authors proposed a rate-splitting multiple access design for a DMA-based architecture under a frequency-selective Lorentzian model to maximize the minimum achievable sum rate across all subcarriers. However, DMA-based wideband ISAC systems under the frequency-selective Lorentzian response model remain largely unexplored. Addressing this gap requires an efficient optimization framework capable of handling the nonlinear and frequency-dependent responses of DMA elements.

Conventional iterative optimization techniques are widely adopted for resource management in DMA-based architectures because of their interpretability and well-established theoretical properties \cite{ID11, ID16, ID19, Rasoul}. For example, the authors in \cite{ID11} maximized the achievable sum rate of a DMA-based multi-user architecture in the near-field region using fractional programming methods. The authors in \cite{ID16} considered a power-efficient DMA-based system and developed a successive convex approximation (SCA)-based method to minimize the transmit power of each DMA-based array. Nevertheless, conventional iterative optimization techniques are often computationally intensive and likely too slow to be implemented within a coherence interval, while also involving several algorithm-specific hyperparameters whose tuning can significantly affect performance and is typically performed manually.

An alternative approach for solving resource management problems in metasurface-based ISAC systems is to use data-driven deep learning methods \cite{ID22, ID23, ID24, ID25, ID25.1}. For example, in \cite{ID23}, the authors proposed an artificial intelligence-based framework for dynamic resource management in ISAC systems, where deep reinforcement learning is employed to optimize beamforming, interference mitigation, and power allocation according to real-time network conditions. In \cite{ID25}, the authors considered a reconfigurable intelligent surface mounted on an unmanned aerial vehicle to improve ISAC coverage and signal quality, with deep reinforcement learning applied to solve a max–min rate optimization problem. Nevertheless, most existing approaches rely on black-box deep neural network (DNN) architectures originally developed for conventional deep learning tasks \cite{ID2}. Such purely data-driven methods generally lack interpretability, and their training usually requires large datasets.

Model-based deep learning, particularly deep unfolding \cite{ID28, ID28.1, ID29, ID29.1}, offers a promising way to alleviate the limitations of both conventional optimization and purely data-driven learning methods. Deep unfolding improves iterative optimization by embedding a fixed number of optimizer iterations into a trainable architecture, where learning is used to refine the update process \cite{ID28.1}. Although recent studies have demonstrated the effectiveness of deep unfolding for rapid hybrid beamforming design in ISAC systems \cite{ID2}, its use in DMA-based wideband ISAC remains unexplored. This gap is particularly important because DMA architectures in wideband systems are more accurately characterized by the frequency-selective Lorentzian model, which leads to nonlinear variations in the responses of DMA elements over the signal bandwidth. Such frequency-dependent behavior, together with the increasing number of optimization variables in emerging systems, requires more sophisticated beamforming designs and makes the convergence of conventional iterative solvers more challenging and highly sensitive to algorithmic hyperparameters. This motivates the use of an unfolded optimization framework for DMA-based wideband ISAC systems, where the model-based update rules are preserved while key optimizer parameters are learned from data.

\subsection{Motivations and Contributions}
Most existing studies on DMA-based ISAC systems adopt the {\it frequency-flat} Lorentzian model \cite{ID13, SM2, RasoulICC}. However, this simplified model fails to accurately characterize the electromagnetic behavior of DMA elements and results in noticeable performance degradation in wideband and ultra-wideband scenarios \cite{Rasoul}. In this paper, we propose a beamforming design for DMA-based wideband ISAC systems under the {\it frequency-selective} Lorentzian model, which captures the frequency-dependent electromagnetic behavior of DMA elements. The resulting beamforming problem is highly challenging due to the strong coupling among the decision variables, namely the resonance frequencies and damping factors of the DMA elements and the digital beamforming matrices. To solve this challenging optimization problem, we develop an alternating projected gradient ascent (PGA)-based optimization framework. Since the frequency-selective Lorentzian model in DMA-based architectures makes the convergence of iterative solvers highly sensitive to algorithmic hyperparameters, we unfold the alternating PGA method into a trainable architecture. The proposed deep-unfolded PGA method preserves the model-based gradient and projection operations of PGA while learning the step sizes from data to improve the objective value and achieve fast convergence. The main contributions are summarized as follows:

\begin{itemize}

    \item We propose a beamforming design for DMA-based wideband ISAC systems. We show that using the frequency-flat approximation rather than the frequency-selective Lorentzian model in DMA-based wideband beam management problems can introduce magnitude errors of several dB and phase deviations on the order of tens of degrees. Motivated by these observations, we use the frequency-selective Lorentzian model to characterize the frequency response of the DMA elements.

        
   \item To address the formulated non-convex optimization problem, we develop an alternating-optimization-based PGA framework that jointly optimizes the digital precoding matrices, resonance frequencies, and damping factors of the DMA elements. We derive closed-form gradients of the objective function with respect to all optimization variables and use them to obtain efficient PGA updates.

    \item To overcome the slow convergence of the PGA optimization framework and improve its performance, we develop a deep-unfolding framework that learns the PGA step sizes from data and treats them as trainable hyperparameters, enabling the proposed method to maximize the objective value within a fixed number of iterations. In particular, the modified PGA is recast as a trainable machine learning architecture, where the step sizes corresponding to the updates of the decision variables are learned. This preserves the interpretability and flexibility of the underlying optimizer while significantly improving convergence speed through a data-driven and adaptive step-size selection mechanism.

    \item We conduct extensive simulations for the considered DMA-based wideband ISAC system. These results show that the proposed deep-unfolded PGA method achieves up to 20-fold faster convergence and improves the objective value by up to 7\% compared with PGA-based benchmarks, while providing about 3-fold and 3.75-fold gains over maximum ratio transmission (MRT) precoding and zero-forcing (ZF) precoding, respectively. Furthermore, the adopted frequency-selective Lorentzian model improves performance by about 20\% over its frequency-flat approximation in the considered scenario.  
    
\end{itemize}
\subsection{Paper Organization and Notations}
The remainder of this paper is organized as follows. Section \ref{system model}  presents the system model, the wideband DMA response model, and the ISAC signal framework. Section \ref{proposed solution} develops the proposed PGA and deep-unfolded PGA-based beamforming schemes for DMA-based wideband ISAC systems. Section \ref{simulations} presents and discusses the simulation results. Finally, Section \ref{conclusion} concludes the paper.

Throughout this paper, scalars, vectors, and matrices are denoted by lowercase, bold lowercase, and bold uppercase letters, respectively. The transpose, Hermitian transpose, and complex conjugate are denoted by $(\cdot)^T$, $(\cdot)^H$, and $(\cdot)^*$, respectively. The Euclidean norm of a vector $\mathbf{x}$ and the magnitude of a scalar $x$ are denoted by $\Vert\mathbf{x}\Vert$ and $\vert x \vert$, respectively. The vectorization operator is denoted by $\mathrm{vec}(\cdot)$. The symbols $\otimes$ and $\odot$ represent the Kronecker and Hadamard products, respectively. The operator $\text{diag}(\mathbf{x})$ denotes a diagonal matrix with the entries of vector $\mathbf{x}$ on its main diagonal. Moreover, $\Re\lbrace\cdot\rbrace$ and $\Im\lbrace\cdot\rbrace$ denote the real and imaginary parts of a complex-valued quantity, respectively. The notation $\mathcal{CN}(\mu,\sigma^2)$ represents a circularly symmetric complex Gaussian random variable with mean $\mu$ and variance $\sigma^2$. Finally, $\mathbf{I}_n$ denotes the $n \times n$ identity matrix, $\mathbb{E}\lbrace\cdot\rbrace$ denotes the expectation operator, $\mathbb{C}$ represents the set of complex numbers, and $[\mathbf{x}]_i$ represents the $i$-th entry of the vector $\mathbf{x}$.
 
\begin{figure}
	\centering
	\includegraphics[scale=0.65]{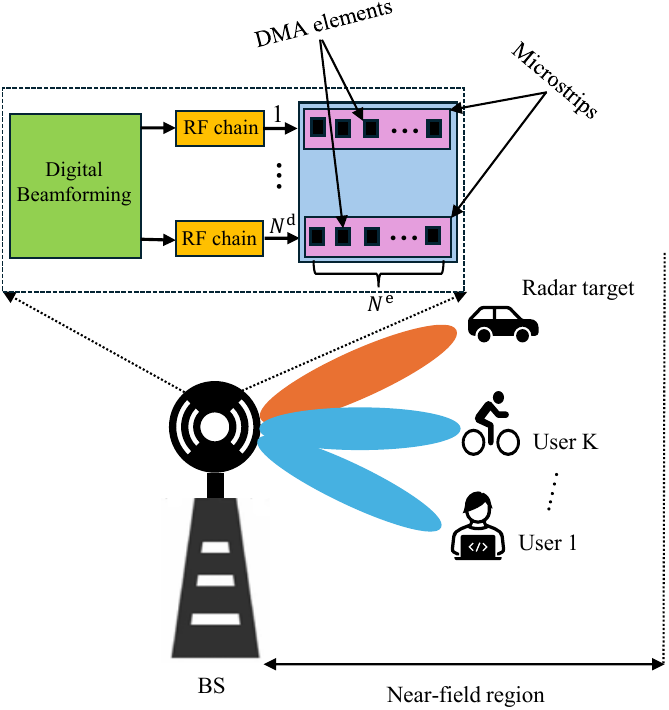}
	\caption{Considered DMA-based ISAC system model.}
    \label{fig_system}
\end{figure}

\section{System model and Problem Formulation}\label{system model}
We consider a DMA-based wideband ISAC system operating in the near-field region. The DMA architecture consists of $N = N^{\mathsf d}\times N^{\mathsf e}$ elements, where $N^{\mathsf d}$ denotes the number of microstrips, and $N^{\mathsf e}$ indicates the number of radiating elements per microstrip, as shown in Fig. \ref{fig_system}. Moreover, the spacing between adjacent microstrips as well as between neighboring radiating elements is set to $d = \frac{\lambda^{\mathsf c}}{2}$, where $\lambda^{\mathsf c}$ represents the wavelength at the central carrier frequency. The BS serves $K$ single-antenna communication users and a radar target. To mitigate the inter-symbol interference inherent in wideband communications, OFDM is employed in the considered system \cite{SM1}. Let $M$ denote the number of OFDM subcarriers. The frequency of the $m$-th subcarrier is expressed as $f_m = f^{\mathsf c} + \frac{B(2m-1-M)}{2M}, \forall m \in \mathcal{M} = \lbrace 1, \dots, M\rbrace$, where $B$ denotes the system bandwidth and $f^{\mathsf c}$ indicates the central carrier frequency \cite{SM1}.

\subsection{DMA Characteristics in Wideband Scenarios}
In a DMA-based architecture, the signals to be radiated by the metamaterial elements are supplied to each waveguide through its input port. The relation between the radiated signals and the signals at the output port of each waveguide is determined by the following two fundamental properties of the DMA architecture \cite{SM4}:

\subsubsection{Microstrip Propagation Model}
In DMA architectures, signals propagating along the microstrip lines experience two primary effects: attenuation and phase shift. The attenuation level is primarily determined by the physical characteristics and geometric dimensions of the microstrip, while the phase shift depends on the wavenumber and the spatial position of the observation point along the waveguide \cite{SM2}. As a result, the complex signal observed at each radiating element can be expressed as follows:
\begin{equation}\label{eq5}
    h_{m,i,l} = e^{-\rho_{i,l}(\alpha_{i,m} + j\beta_{i,m})}, \quad \forall m,i,l,
\end{equation}
where $\rho_{i,l}$ denotes the location of the $l$-th element in the $i$-th microstrip, with $l=1,\dots,N^{\mathsf e}$ and $i=1,\dots,N^{\mathsf d}$. Furthermore, $\alpha_{i,m}$ denotes the attenuation coefficient of the waveguide of the $i$-th microstrip at the $m$-th subcarrier, and $\beta_{i,m}$ represents the phase propagation coefficient of the $i$-th microstrip at the $m$-th subcarrier. We assume that all microstrips are of the same type; $\alpha_{i,m}=\alpha_{m}$ and $\beta_{i,m}=\beta_{m}$. Furthermore, the phase propagation coefficient is computed as $\beta_m = \frac{2\pi}{\lambda_m}\sqrt{e^{\mathsf f}_m}$, where $\lambda_m$ is the wavelength at the $m$-th subcarrier and $e^{\mathsf f}_m$ denotes the effective dielectric coefficient at the $m$-th subcarrier \cite{DMApropagation}. In addition, the microstrip line attenuation coefficient is formulated as $\alpha_m = \alpha_m^{\mathsf d} + \alpha_m^{\mathsf c}$, where $\alpha_m^{\mathsf d}$ and $\alpha_m^{\mathsf c}$ denote the dielectric loss and the conductor loss at the $m$-th subcarrier, respectively  \cite{DMApropagation}. Finally, the DMA propagation matrix at the $m$-th subcarrier ${\mathbf{H}_m\in \mathbb{C}^{N\times N} }$ is formulated as $\mathbf{H}_{m,(i-1)N^{\mathsf e}+l, (i-1)N^{\mathsf e}+l} = h_{m,i,l}$.

\subsubsection{Frequency Response Model}
Each DMA element can be represented as a resonant electrical circuit, with its frequency response described by a Lorentzian function, as established in the literature \cite{SM5, SM6}. The frequency response is expressed as follows:
\begin{equation}\label{eq6}
    q_{m}(f^{\mathsf R}_{i,l}, \kappa_{i,l}) = \frac{F_{i,l} f_m^2}{(f_{i,l}^{\mathsf R})^2 - f_m^2 - j\kappa_{i,l} f_m},
\end{equation}
where $f_{i,l}^{\mathsf R} > 0$ denotes the resonance frequency, $\kappa_{i,l} > 0$ is the damping factor, and $F_{i,l} > 0$ indicates the oscillator strength of the $(i,l)$-th DMA element. The parameters $f_{i,l}^{\mathsf R}$ and $\kappa_{i,l}$ can be independently tuned to control the element response \cite{Rasoul, RESPRAC1, RESPRAC2, RESPRAC3, RESPRAC4, RESPRAC5}. To obtain a more compact formulation of the resonance frequency and the damping factor, we define $\mathbf{f}^{\mathsf R} = \left[f_{1,1}^{\mathsf R}, \dots, f_{N^{\mathsf d},N^{\mathsf e}}^{\mathsf R} \right]$ and $\boldsymbol{\kappa} = \left[\kappa_{1,1}, \dots,  \kappa_{N^{\mathsf d},N^{\mathsf e}}\right]$, respectively. Therefore, the configurable DMA weight matrix at the $m$-th subcarrier $\mathbf{Q}_m \in \mathbb{C}^{N\times N^{\mathsf d}}$ is formulated as follows:
\begin{equation}\label{eq7}
    \mathbf{Q}_{m,(i-1)N^{\mathsf e}+l,n} =
    \begin{cases}
     q_{m}(f^{\mathsf R}_{i,l}, \kappa_{i,l}) & i=n,\\
      0 & i\neq n.
    \end{cases}       
\end{equation}
where $n=1, \dots, N^{\mathsf d}$. In narrowband transmission scenarios, the frequency response of the metamaterial elements in \eqref{eq6} is often approximated as frequency-flat, expressed as $q_{i,l} \in \lbrace\frac{j + e^{j\phi_{i,l}}}{2}| \phi_{i,l} \in [0, 2\pi]\rbrace$. This approximation simplifies the configuration of DMA elements; it is strictly valid only for narrowband signals and becomes inaccurate for wideband and ultra-wideband transmissions \cite{Rasoul}.
\begin{figure}
    \centering
    \begin{minipage}{0.5\textwidth}
        \centering
        \includegraphics[width=\linewidth]{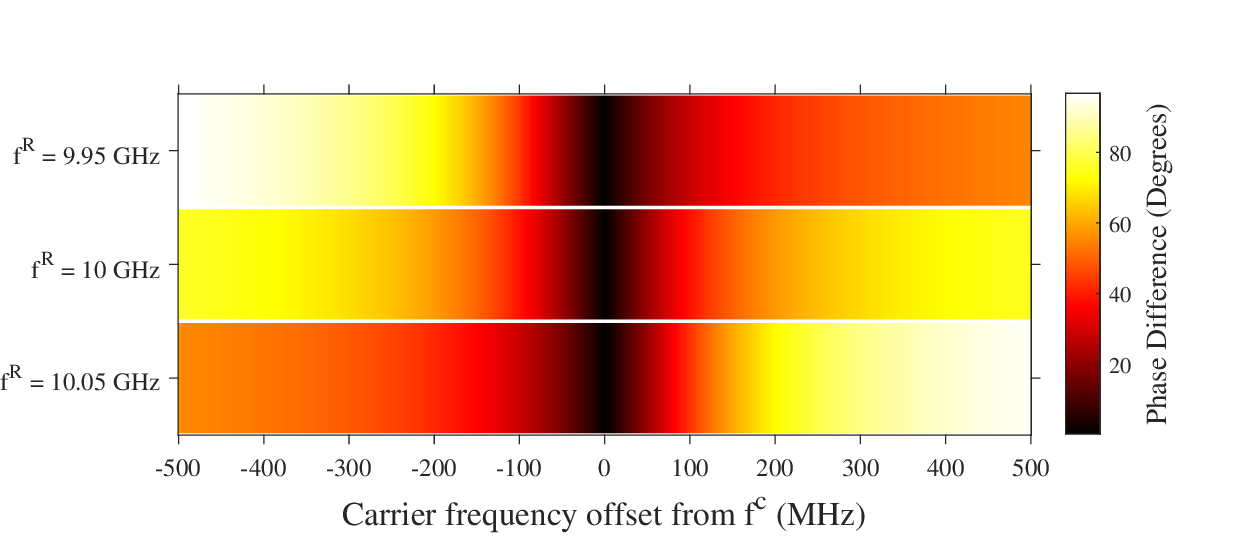}
        \caption{Heatmap of the absolute phase difference between the Lorentzian model and the frequency-flat approximation model.}
        \label{Magnitude}
    \end{minipage}
    \hfill
    \begin{minipage}{0.5\textwidth}
        \centering
        \includegraphics[width=\linewidth]{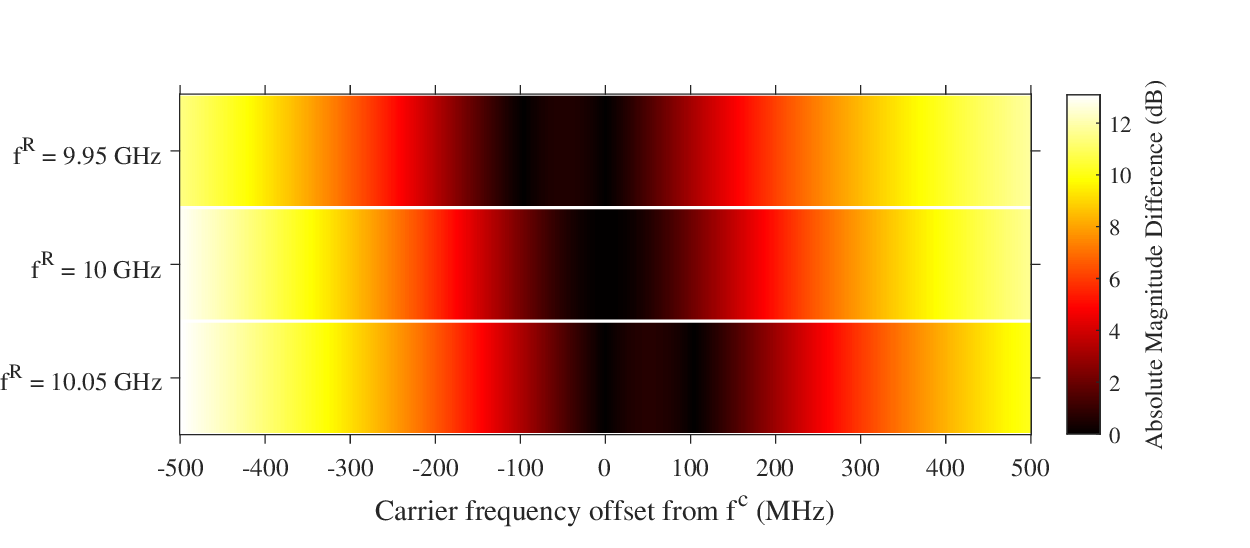}
        \caption{Heatmap of the absolute magnitude difference between the Lorentzian model and the frequency-flat approximation model.} 
        \label{phase}
    \end{minipage} 
\end{figure}

To quantify the modeling error introduced by the frequency-flat approximation, Figs. \ref{Magnitude} and \ref{phase} present the absolute phase and magnitude differences, respectively, between the frequency-selective Lorentzian model and its frequency-flat approximation. Specifically, the figures show the absolute phase and magnitude differences over a $1$ GHz OFDM bandwidth centered at $f^{\mathsf c}=10$ GHz for three resonance frequencies: $f^{\mathsf R}=f^{\mathsf c}$, $f^{\mathsf R}=0.995f^{\mathsf c}$, and $f^{\mathsf R}=1.005f^{\mathsf c}$. The damping factor and the oscillator strength are fixed to $50$ and $1$, respectively \cite{ID15}.

As shown in Figs. \ref{Magnitude} and \ref{phase}, when the resonance frequency of an element is equal to, or close to, the central carrier frequency $f^{\mathsf c}$, the Lorentzian response exhibits pronounced frequency selectivity. In this case, both the magnitude and phase vary significantly across the considered bandwidth. Hence, the frequency-flat approximation fails to capture several dB of gain variation and introduces phase deviations on the order of tens of degrees. Specifically, at a carrier frequency offset of $+250$ MHz, the phase differences for
$f^{\mathsf R} = 9.95$ GHz, $f^{\mathsf R} = 10$ GHz, and $f^{\mathsf R} = 10.05$ GHz are approximately
$45.33^\circ$, $63.15^\circ$, and $79.36^\circ$, respectively.
The corresponding absolute magnitude differences are approximately
$7.37$ dB, $6.69$ dB, and $4.57$ dB. These results confirm that the frequency-flat approximation can introduce substantial amplitude and phase mismatches in wideband systems.

\subsection{Wideband Near-Field Channel Model}
It is assumed that all users and the radar target are located within the near-field region of the BS, i.e., the distances between the users/radar target and the BS are less than the Rayleigh distance, which is computed as $d^{\mathsf F} = \frac{2D^2}{\lambda^{\mathsf c}}$, where $D$ denotes the largest dimension of the antenna aperture. Increasing the number of elements in an antenna array expands its physical size, which in turn enlarges the largest dimension of the antenna aperture, while utilizing higher frequency bands decreases the associated wavelength. This combination of extremely large antenna arrays and high-frequency transmission significantly increases the Rayleigh distance, often reaching several tens of meters. As a result, the traditional far-field approximation, which assumes plane-wave electromagnetic (EM) propagation, becomes invalid at practical operating distances. To properly capture the propagation behavior in this regime, we employ a near-field wireless channel model \cite{SM3}. 

Let $\mathbf{g}^{\mathsf u}_{m,k} \in \mathbb{C}^{N \times 1}$ denote the near-field channel vector from the DMA to user $k$ at subcarrier $m$. Its $(i,l)$-th element is given as: 
\begin{equation}\label{eq1}
    \left[\mathbf{g}^{\mathsf u}_{m,k}\right]_{(i-1)N^{\mathsf e}+l} = \Omega^{\mathsf u}_{m}(\mathbf{p}^{\mathsf t}_{i,l}, \mathbf{p}^{\mathsf u}_{k}) \text{e}^{-j\frac{2\pi f_m}{c}\vert \mathbf{p}^{\mathsf u}_{k} - \mathbf{p}^{\mathsf t}_{i,l}\vert}, \quad \forall m,k,
 \end{equation}
where $\Omega^{\mathsf u}_{m}(\mathbf{p}^{\mathsf t}_{i,l}, \mathbf{p}^{\mathsf u}_{k})$ denotes the channel gain from the $(i,l)$-th DMA element to user $k$ at the $m$-th subcarrier, with $\mathbf{p}^{\mathsf t}_{i,l} = (x_i, y_l, 0)$ and $\mathbf{p}^{\mathsf u}_{k} = (x_k, y_k, z_k)$ being the location of the $l$-th radiating element in the $i$-th microstrip and of the $k$-th user, respectively. The channel gain can be modeled as \cite{SM3}:
\begin{equation}\label{eq2}
     \Omega^{\mathsf u}_{m}(\mathbf{p}^{\mathsf t}_{i,l}, \mathbf{p}^{\mathsf u}_{k}) = \sqrt{T(\Theta_{i,l,k})} \frac{c}{4 \pi f_m \vert \mathbf{p}^{\mathsf u}_{k} - \mathbf{p}^{\mathsf t}_{i,l}\vert},
 \end{equation}
where $T(\Theta_{i,l,k})$ represents the radiation pattern of each radiating element, $\Theta_{i,l,k}=(\theta_{i,l,k}, \varphi_{i,l,k})$ represents the elevation-azimuth pair from the $l$-th element of the $i$-th row of the transmitter to the $k$-th user and $c$ denotes the speed of light. Furthermore,  $T(\Theta_{i,l,k})$ is modeled as follows \cite{SM3}:
\begin{equation}\label{eq3}
  T(\Theta_{i,l,k}) =
    \begin{cases}
      2(b+1)\cos^b(\theta_{i,l,k}), & \theta_{i,l,k}\in[0,\frac{\pi}{2}],\\
      0, & \text{otherwise},  
    \end{cases}       
\end{equation}
where $b$ is the boresight gain, determined by the specific antenna technology. In this study, we focus on the dipole antenna case, for which $b = 2$. The sensing channel can be modeled in a similar manner. Specifically, let
$\mathbf{g}^{\mathsf r}_{m} \in \mathbb{C}^{N \times 1}$ denote the near-field
channel from the DMA to the radar target at subcarrier $m$. The $(i,l)$-th
element of $\mathbf{g}^{\mathsf r}_{m}$ can be modeled following
\eqref{eq1}--\eqref{eq3}. For brevity, the detailed expressions are omitted. 
 
\subsection{Signal Model}
Let $\mathbf{W}^{\mathsf u}_m = \left[\mathbf{w}^{\mathsf u}_{m,1}, \mathbf{w}^{\mathsf u}_{m,2}, \dots, \mathbf{w}^{\mathsf u}_{m,K} \right] \in \mathbb{C}^{N^{\mathsf d}\times K}$ and $\mathbf{W}^{\mathsf r}_m = \left[\mathbf{w}^{\mathsf r}_{m,1}, \mathbf{w}^{\mathsf r}_{m,2}, \dots, \mathbf{w}^{\mathsf r}_{m,N^{\mathsf d}} \right]\in \mathbb{C}^{N^{\mathsf d}\times N^{\mathsf d}}$ represent the digital beamforming matrices for communication and radar  sensing at the $m$-th subcarrier, respectively. Additionally, $\mathbf{s}^{\mathsf r}_m \in \mathbb{C}^{N^{\mathsf d}\times 1}$ and $\mathbf{s}^{\mathsf u}_m\in \mathbb{C}^{K\times 1}$ denote the radar signal and the communication symbols intended for $K$ users, respectively.  Accordingly, the transmit signal radiated by the DMA on the $m$-th subcarrier can be expressed as \cite{SM8}:
\begin{equation}\label{eq8}
    \mathbf{x}_m = \mathbf{H}_m \mathbf{Q}_m\mathbf{W}^{\mathsf r}_m \mathbf{s}^{\mathsf r}_m + \mathbf{H}_m \mathbf{Q}_m\mathbf{W}^{\mathsf u}_m \mathbf{s}^{\mathsf u}_m  .
\end{equation}

Let $\mathbf{W}_m = \left[\mathbf{W}^{\mathsf r}_m \mathbf{W}^{\mathsf u}_m\right] \in \mathbb{C}^{N^{\mathsf d}\times (N^{\mathsf d}+K)}$ denote the joint precoding matrix that maps the radar probing signals and communication symbols to the DMA input at the $m$-th subcarrier. It is assumed that the radar and communication signals are statistically independent, thereby satisfying $\mathbb{E}\left[\mathbf{s}^{\mathsf r}_m(\mathbf{s}^{\mathsf u}_m)^H \right] = \mathbf{0}$. In addition, we assume that the radar signal and the communication symbols satisfy $\mathbb{E}\left[\mathbf{s}_m^{\mathsf r}(\mathbf{s}_m^{\mathsf r})^H\right] = \mathbf{I}_{N^{\mathsf d}}$ and $\mathbb{E}\left[\mathbf{s}_m^{\mathsf u}(\mathbf{s}_m^{\mathsf u})^H\right] = \mathbf{I}_{K}$. Consequently, the effective transmit covariance matrix represents the spatial power distribution and correlation properties of the signals emitted by the DMA, and is formulated as follows \cite{SM2}:
\begin{equation}\label{eq9}
    \mathbf{R}_m = \mathbf{H}_m \mathbf{Q}_m\mathbf{W}_m\mathbf{W}_m^H\mathbf{Q}_m^H\mathbf{H}_m^H.
\end{equation}

Let $\mathbf{G}^{\mathsf r}_m = \zeta_m^{\mathsf r} \mathbf{g}^{\mathsf r}_m(\mathbf{g}^{\mathsf r}_m)^H \in \mathbb{C}^{N\times N}$ denote the end-to-end radar sensing channel at subcarrier $m$, accounting for both the incident and reflected propagation paths associated with the radar target, where $\zeta_m^{\mathsf r}\in\mathbb{C}$ is the target’s complex-valued reflection coefficient. Therefore, the echo signal received from a point target at the $m$-th subcarrier can be expressed as follows:
\begin{equation}\label{eq10}
    \mathbf{y}^{\mathsf r}_m = \mathbf{G}^{\mathsf r}_m \mathbf{x}_m +  \mathbf{n}^{\mathsf r}_m,
\end{equation}
where $\mathbf{n}^{\mathsf r}_m \sim\mathcal{CN}(0,{\tilde{\sigma}}^2_{m}\mathbf{I}_{N})$ denotes the additive white
Gaussian noise at the receiving antennas. The SNR of the radar at the $m$-th subcarrier is formulated as follows \cite{SM2}:
\begin{equation}\label{eq11}
   \gamma^{\mathsf r}_m = \frac{1}{N \tilde{\sigma}_m^2} \text{tr} \left(\mathbf{G}^{\mathsf r}_m \mathbf{R}_m  (\mathbf{G}^{\mathsf r}_m)^H \right).
\end{equation}

The received signal at the $k$-th user over the $m$-th subcarrier can be expressed as follows:
\begin{equation}\label{eq12}
   y_{m,k}^{\mathsf u} = (\mathbf{g}^{\mathsf u}_{m,k})^H(\mathbf{H}_m \mathbf{Q}_m\mathbf{W}^{\mathsf r}_m \mathbf{s}^{\mathsf r}_m + \mathbf{H}_m \mathbf{Q}_m\mathbf{W}^{\mathsf u}_m \mathbf{s}^{\mathsf u}_m)+n_{m,k}^{\mathsf u},
\end{equation}
where $n_{m,k}^{\mathsf u}\sim\mathcal{CN}(0,\sigma^2_{m,k})$ denotes the additive white Gaussian noise for the $k$-th user at the $m$-th subcarrier. Therefore, the SINR of the $k$-th user at the $m$-th subcarrier is formulated as follows \cite{SM2}:
\begin{equation}\label{eq13}
    \gamma_{m,k}^{\mathsf u} = \frac{\vert\mathbf{v}_{m,k}\mathbf{w}^{\mathsf u}_{m,k} \vert^2}{\sum_{i=1,i\neq k}^{K}\vert \mathbf{v}_{m,k} \mathbf{w}_{m,i}^{\mathsf u} \vert^2+ \sum_{j=1}^{N^{\mathsf d}}\vert\mathbf{v}_{m,k} \mathbf{w}_{m,j}^{\mathsf r} \vert^2 + \sigma_{m,k}^2},
\end{equation}
where $\mathbf{v}_{m,k} = (\mathbf{g}^{\mathsf u}_{m,k})^H\mathbf{H}_m \mathbf{Q}_m$. In addition, the system power consumption is formulated as $P = \sum_{m=1}^{M}\Vert \mathbf{H}_m\mathbf{Q}_m\mathbf{W}_{m} \Vert_{\mathcal{F}}^2$.

\subsection{Problem Formulation}
In this paper, we consider a wideband ISAC system employing a DMA-equipped BS, where the resonance frequencies $\mathbf{f}^{\mathsf R}$, damping factors $\boldsymbol{\kappa}$, and digital beamforming matrix $\mathbf{W}_{m}, \forall m$, are jointly optimized to balance communication and sensing performance. Specifically, we maximize a weighted objective that combines the aggregate SINR of communication users and the SNR of the radar target, subject to the total transmit-power budget, DMA hardware constraint, and the quality of service (QoS) requirement of each user. The resulting optimization problem is formulated as follows:
\begin{subequations}\label{eq15}
\begin{align}
\underset{\{\mathbf{f}^{\mathsf R}, \boldsymbol{\kappa}, \mathbf{W}_{m}\}}{\operatorname{maximize}}
&\quad
\xi \sum_{m=1}^{M}\sum_{k=1}^{K} \gamma^{\mathsf u}_{m,k}
+
(1-\xi)\sum_{m=1}^{M} \gamma^{\mathsf r}_{m}
\label{eq15a}\\
\operatorname{subject~to}
&\quad
\sum_{m=1}^{M}
\left\Vert \mathbf{H}_m\mathbf{Q}_m\mathbf{W}_{m} \right\Vert_{\mathcal{F}}^2
\le P^{\mathsf{max}},
\label{eq15b}\\
&\quad
q_{m}(f^{\mathsf R}_{i,l}, \kappa_{i,l}) \in \mathcal{Q}_{m,i,l}, \quad \forall m,i,l, 
\label{eq15c}\\
&\quad
\sum_{m=1}^{M} \gamma_{m,k}^{\mathsf u}
\ge \Gamma_{k}^{\mathsf{min}}, \quad \forall k,
\label{eq15d}
\end{align}
\end{subequations}
 where $\xi \in \left[0,1 \right]$ denotes the weighting parameter that controls the tradeoff between the two objectives. When $\xi$ approaches one, the optimization places more emphasis on improving the aggregate user SINR, whereas smaller values of $\xi$ prioritize enhancing the radar-target SNR. This formulation enables flexible performance adjustment according to network requirements. Furthermore, ${{\mathcal{Q}}_{m,i,l}\triangleq \bigg\lbrace \frac{F_{i,l} f_m^2}{(f_{i,l}^{\rm R})^2 - f_m^2 - j \kappa_{i,l} f_m} \bigg\vert f_{i,l}^{\rm R} >0, \kappa_{i,l}>0 \bigg\rbrace}$. Constraint \eqref{eq15b} restricts the transmit power of the BS to the maximum power budget $P^{\mathsf{max}}$. Constraint \eqref{eq15c} enforces that the frequency response of the configurable DMA weights follows the frequency-selective Lorentzian model.  Finally, constraint \eqref{eq15d} guarantees the QoS requirement of the $k$-th user, where $\Gamma_{k}^{\mathsf {min}}$ denotes the minimum SINR requirement of each user $k$ across all subcarriers.  Problem \eqref{eq15} is non-convex because the design variables $\mathbf{f}^{\mathsf R}$, $\boldsymbol{\kappa}$, and $\mathbf{W}_{m}$ are strongly coupled in the objective function and constraints. To handle the QoS constraint in \eqref{eq15d}, we incorporate it into the objective function as a penalty term. Consequently, problem \eqref{eq15} is reformulated as follows: 
\begin{subequations}\label{eq16}
\begin{align}
\underset{\{\mathbf{f}^{\mathsf R}, \boldsymbol{\kappa}, \mathbf{W}_{m}\}}{\operatorname{maximize}}
&\quad
\xi \sum_{m=1}^{M}\sum_{k=1}^{K} \gamma^{\mathsf u}_{m,k}
+
(1-\xi)\sum_{m=1}^{M} \gamma^{\mathsf r}_{m}
+
\omega \Lambda
\label{eq16a}\\
\operatorname{subject~to}
&\quad
\eqref{eq15b},\ \eqref{eq15c},
\label{eq16b}
\end{align}
\end{subequations}
where $\omega$ denotes the regularization factor and $\Lambda = \sum_{k=1}^{K} \ln(\sum_{m=1}^{M}\gamma_{m,k}^{\mathsf u} - \Gamma_{k}^{\mathsf {min}})$. Here, we treat $\omega$ as a given hyperparameter. Problem \eqref{eq16} is still non-convex and challenging to solve.

\section{Proposed Solution}\label{proposed solution}
To solve problem \eqref{eq16}, we develop a PGA-based optimization framework and subsequently unfold it into a trainable architecture. In Section \ref{sectionIIIA}, we derive closed-form gradients of the objective function in \eqref{eq16} with respect to all optimization variables $\lbrace \mathbf{f}^{\mathsf R}, \boldsymbol{\kappa}, \mathbf{W}_{m} \rbrace$. These gradients enable efficient updates of the decision variables for maximizing the objective function in \eqref{eq16}. In Section \ref{sectionIIIB}, we propose a deep-unfolded PGA algorithm that leverages data-driven learning to accelerate convergence and improve the performance for the considered non-convex problem. Finally, we provide a detailed computational complexity analysis of the proposed method in Section \ref{sectionIIIC}. 

\subsection{PGA Optimization Framework}\label{sectionIIIA}
 We develop an alternating PGA-based approach to solve problem \eqref{eq16}. At each iteration, one block of variables is updated while the remaining variables are fixed. Specifically, the algorithm alternately updates $\mathbf{f}^{\mathsf R}$, $\boldsymbol{\kappa}$, and $\mathbf{W}_{m}$. Each update requires the gradient of the objective function with respect to the corresponding variable block, followed by projection onto the feasible set defined by constraint \eqref{eq15b} through a normalization process. We first present the PGA framework and then separately derive the closed-form gradients required for updating each decision variable, as follows:

\subsubsection{Resonance frequency optimization}
Let $\gamma^{\mathsf u, \mathsf {tot}} = \sum_{m=1}^{M}\sum_{k=1}^{K} \gamma^{\mathsf u}_{m,k}$ and $\gamma^{\mathsf r, \mathsf {tot}} = \sum_{m=1}^{M} \gamma^{\mathsf r}_{m}$. Specifically, for a fixed $\mathbf{W}_m$ and $\boldsymbol{\kappa}$, $\mathbf{f}^{\mathsf R}$ can be updated at iteration $i+1$ as follows:
\begin{align}
(\mathbf{f}^{\mathsf R})^{(i+1)}
&= (\mathbf{f}^{\mathsf R})^{(i)} + \mu_{\mathbf{f}^{\mathsf R}}^{(i)}
\Big(
\xi \nabla_{\mathbf{f}^{\mathsf R}} \gamma^{\mathsf u,\mathsf{tot}}
+ (1-\xi)\nabla_{\mathbf{f}^{\mathsf R}} \gamma^{\mathsf r,\mathsf{tot}}
\nonumber\\
&\hspace{2.6cm}
+ \omega \nabla_{\mathbf{f}^{\mathsf R}} \Lambda
\Big)
\Big|_{\mathbf{f}^{\mathsf R}=(\mathbf{f}^{\mathsf R})^{(i)}},
\label{eq19}
\end{align}
where $\mu_{\mathbf{f}^{\mathsf R}}^{(i)}$  denotes the step size at the $i$-th PGA iteration for updating $\mathbf{f}^{\mathsf R}$ and $\nabla_{\mathbf{X}} f$ is the gradient of a scalar-valued function $f$ with respect to a complex matrix $\mathbf{X}$. The update rule in \eqref{eq19} requires closed-form expressions for the gradients of $\gamma^{\mathsf u, \mathsf {tot}}$, $\gamma^{\mathsf r, \mathsf {tot}}$, and $\Lambda$ with respect to $\mathbf{f}^{\mathsf R}$. The following theorem provides the closed-form gradients required for updating $\mathbf{f}^{\rm R}$.

{\it Theorem 1:} The gradients of $\gamma^{\mathsf u, \mathsf {tot}}$, $\gamma^{\mathsf r, \mathsf {tot}}$, and $\Lambda$ with respect to $\mathbf{f}^{\mathsf R}$ are given by

\begin{equation}\label{eq29}
    \nabla_{\mathbf{f}^{\mathsf R}}\gamma^{\mathsf u,\mathsf {tot}} = 2 \Re\Big\lbrace \!\! \sum_{m=1}^{M} \sum_{k=1}^{K} \mathbf{b}_m \!\odot\!\! \left(\!\frac{\hat{\mathbf{u}}_{m,k} A_{m,k} - D_{m,k} \hat{\mathbf{t}}_{m,k}}{A_{m,k}^2}\right)^* \Big\rbrace,
\end{equation}

\begin{equation}\label{eq30}
    \nabla_{\mathbf{f}^{\mathsf R}}\gamma^{\mathsf r,\mathsf {tot}} =  \!\Re\Big\lbrace \!\sum_{m=1}^{M} \!\!\frac{2}{N\tilde{\sigma}_m^2} \left(\mathbf{b}_m \!\!\odot\! (\left[(\mathbf{B}_m\otimes \mathbf{I}_{N^{\mathsf e}})\!\odot\!\mathbf{E}_m\right]\hat{\mathbf{q}}_m)^*\right)\!\!\!\Big\rbrace,
\end{equation}

\begin{equation}\label{eq31}
    \nabla_{\mathbf{f}^{\mathsf R}}\Lambda = \sum_{m=1}^{M}\sum_{k=1}^{K}\frac{1}{\sum_{\ell=1}^{M}\gamma_{\ell,k}^{\mathsf u}-\Gamma_{k}^{\mathsf {min}}}\nabla_{\mathbf{f}^{\mathsf R}}
    \gamma_{m,k}^{\mathsf u}, 
\end{equation}
where $\mathbf{b}_m = \left[b_{m,1,1}, \dots, b_{m,N^{\mathsf d},N^{\mathsf e}} \right]^T\in \mathbb{C}^{N \times 1}$, in which $b_{m,i,l}=-\frac{2f^{\mathsf R}_{i,l}(F_{i,l} f_m^2)}{\left((f_{i,l}^{\mathsf R})^2 - f_m^2 - j\kappa_{i,l} f_m\right)^2}$, and $A_{m,k} = \sum_{i=1,i\neq k}^{K}\vert \mathbf{v}_{m,k} \mathbf{w}_{m,i}^{\mathsf u} \vert^2+ \sum_{j=1}^{N^{\mathsf d}}\vert\mathbf{v}_{m,k} \mathbf{w}_{m,j}^{\mathsf r} \vert^2 + \sigma_{m,k}^2$. To simplify the presentation, let $\mathcal{I}_m$ denote the indices of the nonzero entries of $\mathbf{q}_m=\text{vec}(\mathbf{Q}_m)\in \mathbb{C}^{N N^{\mathsf d}\times1}$. For any vector $\mathbf{x}_m$, its modified version $\hat{\mathbf{x}}_m$ is obtained by retaining only the entries indexed by $\mathcal{I}_m$. 
Therefore, $\hat{\mathbf{q}}_m\in \mathbb{C}^{N\times1}$ is the modified version of $\mathbf{q}_m$, and $\hat{\mathbf{u}}_{m,k}\in\mathbb{C}^{N\times 1}$ denotes the modified version of $\mathbf{u}_{m,k}=(\mathbf{z}_{m,k,k}^{\mathsf u})^{H}\mathbf{z}_{m,k,k}^{\mathsf u}\mathbf{q}_m$, where $\mathbf{z}_{m,k,k}^{\mathsf u}=(\mathbf{w}_{m,k}^{\mathsf u})^{T}\otimes(\mathbf{g}_{m,k}^{\mathsf u})^{H}\mathbf{H}_m\in\mathbb{C}^{1\times NN^{\mathsf d}}$. Furthermore, $\hat{\mathbf{t}}_{m,k}\in \mathbb{C}^{N \times 1}$ is the modified version of  $\mathbf{t}_{m,k} = \sum_{i=1,i\neq k}^{K}(\mathbf{z}_{m,i,k}^{\mathsf u})^H\mathbf{z}_{m,i,k}^{\mathsf u}\mathbf{q}_m+ \sum_{j=1}^{N^{\mathsf d}}(\mathbf{z}_{m,j,k}^{\mathsf r})^H\mathbf{z}_{m,j,k}^{\mathsf r}\mathbf{q}_m$, in which $\mathbf{z}^{\mathsf r}_{m,j,k} = (\mathbf{w}^{\mathsf r}_{m,j})^T \otimes (\mathbf{g}^{\mathsf u}_{m,k})^H\mathbf{H}_m\in \mathbb{C}^{1 \times N N^{\mathsf d}}$. In addition, $D_{m,k} = \vert \hat{\mathbf{z}}_{m,k,k}^{\mathsf u}\hat{\mathbf{q}}_m \vert^2$, in which $\hat{\mathbf{z}}_{m,k,k}$ denotes the modified version of ${\mathbf{z}}_{m,k,k}$. Finally, $\mathbf{E}_m = \mathbf{H}_m^H(\mathbf{G}_m^{\mathsf r})^H\mathbf{G}_m^{\mathsf r}\mathbf{H}_m\in \mathbb{C}^{N \times N }$ and $\mathbf{B}_m = \mathbf{W}_m \mathbf{W}^H_m\in \mathbb{C}^{N^{\mathsf d} \times N^{\mathsf d}}$. 

{Proof:} See Appendix \ref{AppendixA}.

\subsubsection{Damping factor optimization}
Given $\mathbf{W}_{m}$ and  $\mathbf{f}^{\mathsf R}$, $\boldsymbol{\kappa}$ can be updated at iteration $i+1$ as follows:
\begin{align}
\boldsymbol{\kappa}^{(i+1)}
&=\boldsymbol{\kappa}^{(i)} + \mu_{\boldsymbol{\kappa}}^{(i)}
\Big(
\xi \nabla_{\boldsymbol{\kappa}} \gamma^{\mathsf u,\mathsf{tot}}
+ (1-\xi)\nabla_{\boldsymbol{\kappa}} \gamma^{\mathsf r,\mathsf{tot}}
\nonumber\\
&\hspace{2.6cm}
+ \omega \nabla_{\boldsymbol{\kappa}} \Lambda
\Big)
\Big|_{\boldsymbol{\kappa}=\boldsymbol{\kappa}^{(i)}},
\label{eq21}
\end{align}
where $\mu_{\boldsymbol{\kappa}}^{(i)}$  denotes the step size at the $i$-th PGA iteration for updating $\boldsymbol{\kappa}$. Implementing \eqref{eq21} requires $\nabla_{\boldsymbol{\kappa}} \gamma^{\mathsf u,\mathsf{tot}}$, $\nabla_{\boldsymbol{\kappa}} \gamma^{\mathsf r,\mathsf{tot}}$, and $\nabla_{\boldsymbol{\kappa}} \Lambda$. The following theorem provides the closed-form gradients required
for updating $\boldsymbol{\kappa}$.

{\it Theorem 2:} The gradients of $\gamma^{\mathsf u, \mathsf {tot}}$, $\gamma^{\mathsf r, \mathsf {tot}}$, and $\Lambda$ with respect to $\boldsymbol{\kappa}$ are given by
\begin{equation}\label{eq32}
    \nabla_{\boldsymbol{\kappa}}\gamma^{\mathsf u,\mathsf {tot}} = 2 \Re\Big\lbrace \sum_{m=1}^{M} \sum_{k=1}^{K} \mathbf{p}_m \odot \left(\frac{\hat{\mathbf{u}}_{m,k} A_{m,k} - D_{m,k} \hat{\mathbf{t}}_{m,k}}{A_{m,k}^2}\right)^* \Big\rbrace,
\end{equation}

\begin{equation}\label{eq33}
    \nabla_{\boldsymbol{\kappa}}\gamma^{\mathsf r,\mathsf {tot}} \!= \!  \Re\Big\lbrace\!\!\sum_{m=1}^{M}\!\!\frac{2}{N\tilde{\sigma}_m^2}\left(\mathbf{p}_m \!\!\odot\!\! \Big(\!\!\left[(\mathbf{B}_m\!\otimes \!\mathbf{I}_{N^{\mathsf e}})\!\odot\!\mathbf{E}_m\right]\hat{\mathbf{q}}_m\!\Big)^*\right)\!\!\!\Big\rbrace,
\end{equation}

\begin{equation}\label{eq34}
    \nabla_{\boldsymbol{\kappa}}\Lambda = \sum_{m=1}^{M}\sum_{k=1}^{K}\frac{1}{\sum_{\ell=1}^{M}\gamma_{\ell,k}^{\mathsf u}-\Gamma_{k}^{\mathsf {min}}}\nabla_{\boldsymbol{\kappa}}
    \gamma_{m,k}^{\mathsf u},
\end{equation}
where $\mathbf{p}_m = \left[p_{m,1,1}, \dots, p_{m,N^{\mathsf d},N^{\mathsf e}} \right]^T\in \mathbb{C}^{N \times 1}$, in which $p_{m,i,l} = \frac{jF_{i,l} f_m^3}{\left((f_{i,l}^{\mathsf R})^2 - f_m^2 - j\kappa_{i,l} f_m\right)^2}$.

{Proof:} See Appendix \ref{AppendixB}.

Next, $\mathbf{Q}_m^{(i+1)}$ is constructed from updated values $(\mathbf{f}^{\mathsf R})^{(i+1)}$ and $\boldsymbol{\kappa}^{(i+1)}$ according to \eqref{eq6} and \eqref{eq7}. 

\subsubsection{Digital beamforming optimization}
For a fixed $\mathbf{f}^{\mathsf R}$ and $\boldsymbol{\kappa}$, $\mathbf{W}_{m}$ can be updated at iteration $i+1$ using a projected gradient ascent step as follows:
\begin{align}
\mathbf{W}_m^{(i+1)}
&= \mathbf{W}_m^{(i)} + \mu_{\mathbf{W}}^{(i)}
\Big(
\xi \nabla_{\mathbf{W}_m} \gamma^{\mathsf u,\mathsf{tot}}
+ (1-\xi)\nabla_{\mathbf{W}_m} \gamma^{\mathsf r,\mathsf{tot}}
\nonumber\\
&\hspace{2.7cm}
+ \omega \nabla_{\mathbf{W}_m} \Lambda
\Big)
\Big|_{\mathbf{W}_m = \mathbf{W}_m^{(i)}},
\label{eq17}
\end{align}
\begin{equation}\label{eq18}
    \mathbf{W}_m^{(i+1)} = \sqrt{\frac{ P^{\mathsf{max}}}{\sum_{m=1}^{M} \Vert \mathbf{H}_m\mathbf{Q}_m^{(i+1)}\mathbf{W}_{m}^{(i+1)} \Vert_{\mathcal{F}}^2}} \mathbf{W}_m^{(i+1)},
\end{equation}
where $\mu_{\mathbf{W}}^{(i)}$  denotes the step size at the $i$-th PGA iteration for updating $\mathbf{W}_m$. This update rule requires the closed-form gradients of $\gamma^{\mathsf u, \mathsf {tot}}$, $\gamma^{\mathsf r, \mathsf {tot}}$, and $\Lambda$ with respect to $\mathbf{W}_{m}$. The following theorem provides the closed-form gradients required
for updating $\mathbf{W}_m$.

{\it Theorem 3:} The gradients of $\gamma^{\mathsf u, \mathsf {tot}}$, $\gamma^{\mathsf r, \mathsf {tot}}$, and $\Lambda$ with respect to $\mathbf{W}_{m}$ are formulated as follows, respectively:
\begin{align}
\nabla_{\mathbf{w}_{m,k}^{\mathsf u}}
\gamma^{\mathsf u,\mathsf{tot}}
&=
\frac{
2\mathbf{v}_{m,k}^{H}\mathbf{v}_{m,k}\mathbf{w}_{m,k}^{\mathsf u}
}{
A_{m,k}
}
\nonumber\\
&\quad
-
2\sum_{\substack{j=1\\ j\neq k}}^{K}
\frac{
\left|\mathbf{v}_{m,j}\mathbf{w}_{m,j}^{\mathsf u}\right|^2
}{
A_{m,j}^{2}
}
\mathbf{v}_{m,j}^{H}\mathbf{v}_{m,j}
\mathbf{w}_{m,k}^{\mathsf u},
\label{eq23}\\[1mm]
\nabla_{\mathbf{w}_{m,j}^{\mathsf r}}
\gamma^{\mathsf u,\mathsf{tot}}
&=
-2\sum_{k=1}^{K}
\frac{
\left|\mathbf{v}_{m,k}\mathbf{w}_{m,k}^{\mathsf u}\right|^2
}{
A_{m,k}^{2}
}
\mathbf{v}_{m,k}^{H}\mathbf{v}_{m,k}
\mathbf{w}_{m,j}^{\mathsf r},
\label{eq24}\\[1mm]
\nabla_{\mathbf{W}_m}
\gamma^{\mathsf r,\mathsf{tot}}
&=
\frac{2}{N\tilde{\sigma}_m^2}
\mathbf{Q}_m^{H}\mathbf{H}_m^{H}
(\mathbf{G}_m^{\mathsf r})^{H}\mathbf{G}_m^{\mathsf r}
\mathbf{H}_m\mathbf{Q}_m\mathbf{W}_m,
\label{eq25}\\[1mm]
\nabla_{\mathbf{W}_m}\Lambda
&=
\sum_{k=1}^{K}
\frac{
\nabla_{\mathbf{W}_m}\gamma_{m,k}^{\mathsf u}
}{
\sum_{\ell=1}^{M}\gamma_{\ell,k}^{\mathsf u}
-
\Gamma_{k}^{\mathsf{min}}
}.
\label{eq26}
\end{align}
{Proof:} See Appendix \ref{AppendixC}.

Using the derived gradients, the update rules  \eqref{eq19}, \eqref{eq21}, and \eqref{eq17}  can be directly applied to compute $\mathbf{f}^{\mathsf R}$, $\boldsymbol{\kappa}$, and $\mathbf{W}_m$, respectively. However, the conventional PGA optimization may suffer from slow or unstable convergence, especially when fixed step sizes are used to update the coupled variables $\mathbf{W}_m$, $\mathbf{f}^{\mathsf R}$, and $\boldsymbol{\kappa}$. This motivates us to use a deep unfolding framework, where the PGA iterations are unrolled into a trainable structure to improve convergence within fewer iterations.
Furthermore, it is important to note that our goal is to achieve fast convergence to solutions for $\mathbf{f}^{\mathsf R}$, $\boldsymbol{\kappa}$, and $\mathbf{W}_m$, ideally within a small number of iterations. Consequently, the step sizes $\lbrace \mu_{\mathbf{f}^{\mathsf R}}^{(i)}, \mu_{\boldsymbol{\kappa}}^{(i)}, \mu_{\mathbf{W}}^{(i)} \rbrace$ in \eqref{eq19}, \eqref{eq21}, and \eqref{eq17} play a crucial role in the performance of the PGA method, and determining them is nontrivial. Although line search and backtracking can adapt the step sizes during runtime, it requires multiple objective evaluations at each iteration, which leads to additional overhead. To avoid this overhead, we unfold the PGA iterations into a trainable architecture and learn the step-size parameters from data (Section \ref{sectionIIIB}).

\subsection{Proposed Deep-Unfolded PGA Model}\label{sectionIIIB}
In this section, we construct the unfolded framework to preserve both the interpretability and flexibility of PGA by strictly following the original PGA operations outlined in Section \ref{sectionIIIA}. Specifically, each layer of the unfolded architecture corresponds to one PGA iteration and performs the same sequence of operations as the model-based algorithm, including gradient evaluation, variable updating, and projection onto the feasible set. In this way, the unfolded model retains the structure of the original optimization procedure while enabling data-driven performance enhancement. We further incorporate the algorithm hyperparameters, specifically the step sizes $\{ \mu^{(i)}_{\mathbf{f}^{\mathsf R}}, \mu^{(i)}_{\boldsymbol{\kappa}}, \mu^{(i)}_{\mathbf{W}} \}$, as trainable parameters within the unfolded model \cite{ID2}. By learning these step sizes from data, the proposed framework can adapt the update trajectory to the considered system model, thereby accelerating convergence and improving the achieved objective value under a fixed number of iterations. The model structure and training procedure are detailed as follows:
\begin{figure*}
	\centering
	\includegraphics[scale=0.55]{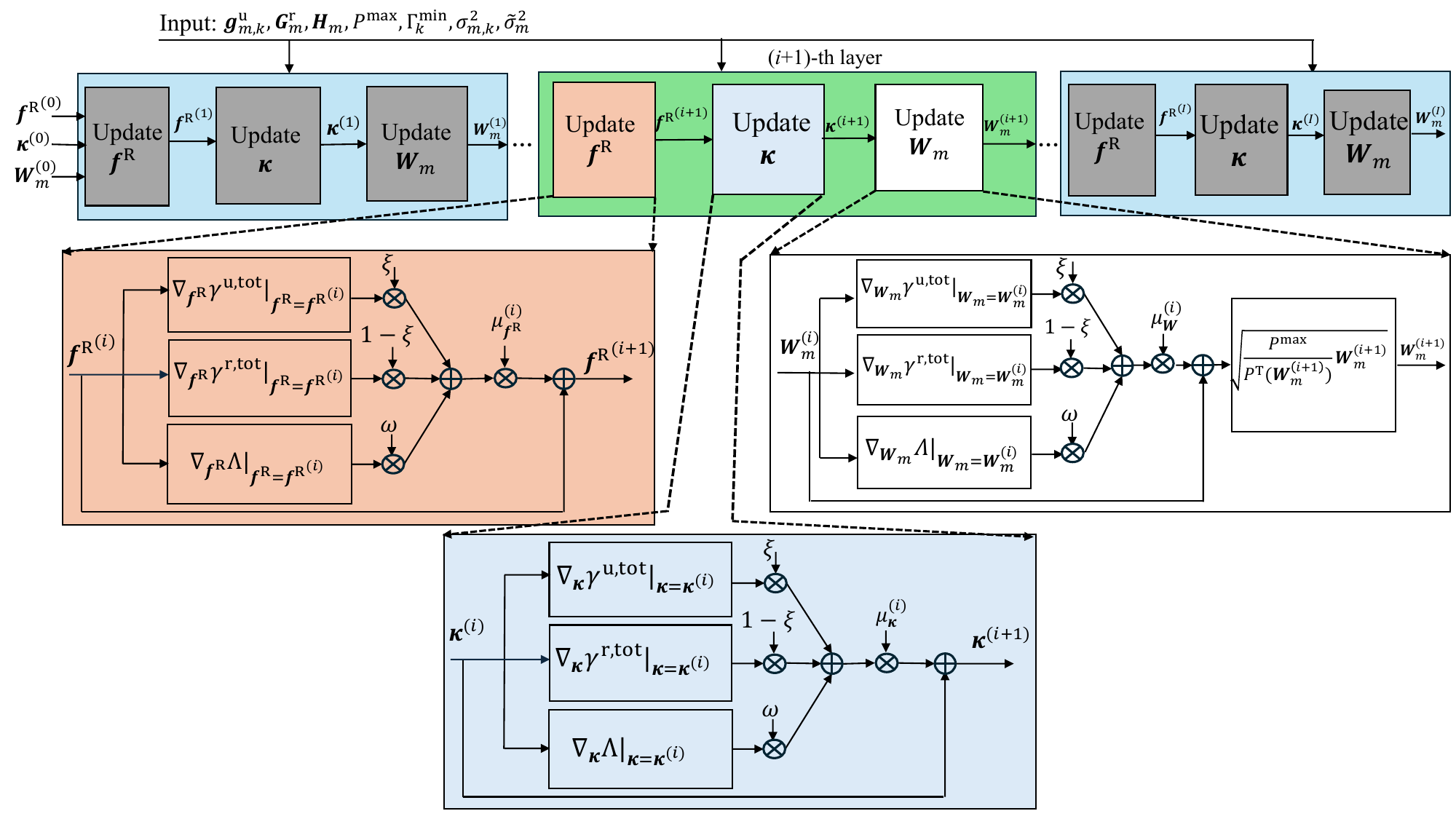}
	\caption{Illustration of the proposed unfolded PGA model.}
    \label{fig_deep}
\end{figure*}

\subsubsection{Model structure}
The proposed $I$-layer deep-unfolded PGA-based DNN maps each iteration of the PGA algorithm onto a corresponding network layer. The objective of this model is to produce feasible design variables $\lbrace \mathbf{f}^{\mathsf R}, \boldsymbol{\kappa}, \mathbf{W}_m \rbrace$ that achieve high communication and sensing performance. Specifically, the model maximizes the following objective function:
\begin{equation}
f(\mathbf{f}^{\mathsf R},\boldsymbol{\kappa}, \mathbf{W}_m)=
\!\xi \sum_{m=1}^{M}\!\sum_{k=1}^{K}\!\gamma^{\mathsf u}_{m,k}
\!+\!
(1\!-\!\xi)\!\sum_{m=1}^{M}\gamma^{\mathsf r}_{m}
+
\omega\Lambda.
\end{equation}

In the unfolding framework, each iteration of the PGA procedure is mapped to a layer in the deep-unfolded PGA model. Accordingly, we continue to use the subscript $i$ to denote the layers when referring to the deep-unfolded PGA model.
Moreover,  we denote $\boldsymbol{\mu}_{\mathbf{f}^{\mathsf R}} = \lbrace \mu_{\mathbf{f}^{\mathsf R}}^{(i)}\rbrace_{i=0}^{I-1}$, $\boldsymbol{\mu}_{\boldsymbol{\kappa}} = \lbrace \mu_{\boldsymbol{\kappa}}^{(i)}\rbrace_{i=0}^{I-1}$, and $\boldsymbol{\mu}_{\mathbf{W}} = \lbrace \mu_{\mathbf{W}}^{(i)}\rbrace_{i=0}^{I-1}$ corresponding to the updates of $\mathbf{f}^{\mathsf R}$, $\boldsymbol{\kappa}$, and $\mathbf{W}_m$, respectively, for ease of exposition. 

The deep-unfolded PGA model, depicted in Fig. \ref{fig_deep}, implements the update procedures described in \eqref{eq19}-\eqref{eq18}. It takes as inputs an initial set of variables $\lbrace (\mathbf{f}^{\mathsf R})^{(0)}, \boldsymbol{\kappa}^{(0)}, \mathbf{W}_m^{(0)}\rbrace$, the channel matrices of the users and the radar $\mathbf{g}_{m,k}^{\mathsf u}$, $\mathbf{g}_{m}^{\mathsf r}$, the DMA propagation matrix $\mathbf{H}_m$, the power budget at the BS $P^{\mathsf \max}$, and the noise variance $\sigma^2_{m,k}$ for each user and the radar at the $m$-th subcarrier, and produces as outputs the updated variables $\lbrace (\mathbf{f}^{\mathsf R})^{(i)}, \boldsymbol{\kappa}^{(i)}, \mathbf{W}_m^{(i)} \rbrace$ across the outer layers $i=1, \dots, I$. 

Each layer computes the gradients specified in \eqref{eq19}, \eqref{eq21}, and \eqref{eq17}, followed by the projections defined in \eqref{eq18}. The DMA configurable weight matrix $\mathbf{Q}_m$ at the $m$-th subcarrier is not directly learned; rather, it is reconstructed at each layer from the updated values of $\mathbf{f}^{\mathsf R}$ and $\boldsymbol{\kappa}$ according to the Lorentzian response model in \eqref{eq6} and the DMA mapping in \eqref{eq7}.

\begin{algorithm}[t!]
\caption{Proposed deep-unfolded PGA-based algorithm for solving \eqref{eq16}.}
\label{Algorithm1}
\begin{algorithmic}[1]

\Statex \textbf{Input:} 
$\mathbf{g}_{m,k}^{\mathsf u}, \forall m,k$, 
$\mathbf{G}_m^{\mathsf r}, \forall m$, 
$\Gamma_k^{\mathsf{min}}$, 
$P^{\mathsf{max}}$, $\xi$, and trained step sizes
$\{\boldsymbol{\mu}_{\mathbf{f}^{\mathsf R}},\boldsymbol{\mu}_{\boldsymbol{\kappa}},\boldsymbol{\mu}_{\mathbf{W}}\}$ obtained from Section \ref{sectionIIIB}.

\Statex \textbf{Output:} 
$\{\mathbf{f}^{\mathsf R},\boldsymbol{\kappa}, \mathbf{W}_m\}$.

\Statex \textbf{Initialization:} 
Generate
$\{\mathbf{W}_m^{(0)},\mathbf{f}^{\mathsf R(0)},
\boldsymbol{\kappa}^{(0)}\}$.

\State $i \gets 0$.

\Repeat

\State Update $\mathbf{f}^{\mathsf R}$ using \eqref{eq19}.

\State Update $\boldsymbol{\kappa}$ using \eqref{eq21}.

\State Construct $\mathbf{Q}_m$, $\forall m$, using \eqref{eq6} and \eqref{eq7}.

\State Update and project $\mathbf{W}_m$, $\forall m$, using \eqref{eq17} and \eqref{eq18}.

\State $i \gets i+1$.
\Until \text{Convergence}.

\end{algorithmic}
\end{algorithm}

\subsubsection{Training the model}
Since the unfolded architecture is derived from the optimization problem in \eqref{eq16}, its training is aimed at maximizing the objective function $f(\mathbf{f}^{\mathsf R}, \boldsymbol{\kappa}, \mathbf{W}_m)$. Consequently, the loss function is defined as follows:
\begin{equation}\label{eq35}
    \mathcal{L}(\boldsymbol{\mu}_{\mathbf{f}^{\mathsf R}},\boldsymbol{\mu}_{\boldsymbol{\kappa}},\boldsymbol{\mu}_{\mathbf{W}}) = -f((\mathbf{f}^{\mathsf R})^{(I)}, \boldsymbol{\kappa}^{(I)}, \mathbf{W}_m^{(I)}).
\end{equation}
The loss function $\mathcal{L}(\boldsymbol{\mu}_{\mathbf{f}^{\mathsf R}},\boldsymbol{\mu}_{\boldsymbol{\kappa}},\boldsymbol{\mu}_{\mathbf{W}})$ allows the model to be trained in an unsupervised manner. Specifically, the dataset consists of multiple channel realizations, and the learned hyperparameters are optimized for a range of user SINRs and radar-target SNRs by appropriately setting the noise power to $\sigma^2_{m,k} = 1$ and $\tilde{\sigma}^2_{m} = 1, \forall m,k$. Our focus is on the moderate-to-high SNR regime, which is typically necessary for effective performance in both communication and radar sensing applications \cite{PS1}. Accordingly, we set  $\xi$ and $\omega$ to balance communication performance, sensing performance, and user QoS in the moderate-to-high SINR regime, and treat them as fixed hyperparameters during training of the unfolded model. The loss function $ \mathcal{L}(\boldsymbol{\mu}_{\mathbf{f}^{\mathsf R}},\boldsymbol{\mu}_{\boldsymbol{\kappa}},\boldsymbol{\mu}_{\mathbf{W}})$ depends on the step sizes $\lbrace\boldsymbol{\mu}_{\mathbf{f}^{\mathsf R}},\boldsymbol{\mu}_{\boldsymbol{\kappa}},\boldsymbol{\mu}_{\mathbf{W}}\rbrace$, as the variables $\lbrace (\mathbf{f}^{\mathsf R})^{(I)}, \boldsymbol{\kappa}^{(I)}, \mathbf{W}_m^{(I)} \rbrace$ are functions of $\lbrace (\mathbf{f}^{\mathsf R})^{(i)} \rbrace_{i=0}^{I-1}$, $\lbrace\boldsymbol{\kappa}^{(i)}\rbrace_{i=0}^{I-1}$, $\lbrace \mathbf{W}_m^{(i)} \rbrace_{i=0}^{I-1}$, and $\lbrace\boldsymbol{\mu}_{\mathbf{f}^{\mathsf R}},\boldsymbol{\mu}_{\boldsymbol{\kappa}},\boldsymbol{\mu}_{\mathbf{W}}\rbrace$. The deep-unfolded PGA model is trained to learn the optimal values of $\lbrace\boldsymbol{\mu}_{\mathbf{f}^{\mathsf R}},\boldsymbol{\mu}_{\boldsymbol{\kappa}},\boldsymbol{\mu}_{\mathbf{W}}\rbrace$ that yield the best tradeoff within $I$ iterations. The proposed deep-unfolded PGA-based algorithm for the DMA-based wideband ISAC system is summarized in Algorithm~\ref{Algorithm1}.

\subsection{Computational Complexity}\label{sectionIIIC}
The proposed algorithm iteratively updates $\mathbf{f}^{\mathsf R}$, $\boldsymbol{\kappa}$, and $\mathbf{W}_m$ to solve problem \eqref{eq16} using the deep-unfolded PGA method. The main computational complexity of Algorithm 1 arises from the computation of the gradients in \eqref{eq23}, \eqref{eq24}, \eqref{eq25}, \eqref{eq29}, \eqref{eq30}, \eqref{eq32}, and \eqref{eq33}. A detailed computational complexity analysis of the proposed Algorithm 1 is provided in Table \ref{tab:multiplications}.
\begin{table}[t]
\centering
\caption{Number of multiplications and additions required in Algorithm~1.}
\label{tab:multiplications}
\renewcommand{\arraystretch}{1.25}
\setlength{\tabcolsep}{3pt}
\begin{tabular}{|c|c|}
\hline
\textbf{Task} & \textbf{No. of multiplications and additions} \\
\hline
\eqref{eq29} &
$3(N^{\mathsf d})^2N+KN^{\mathsf d}N+K^2N^{\mathsf d}N
+K^2N+3KN^2+KN$ \\
\hline
\eqref{eq30} &
$N^{\mathsf d}(N^{\mathsf d}+K)^2+3N^2$ \\
\hline
\eqref{eq32} &
$3(N^{\mathsf d})^2N+KN^{\mathsf d}N+K^2N^{\mathsf d}N
+K^2N+3KN^2+KN$ \\
\hline
\eqref{eq33} &
$N^{\mathsf d}(N^{\mathsf d}+K)^2+3N^2$ \\
\hline
\eqref{eq23} & $K (N^{\mathsf d})^2+K^2N^{\mathsf d} + KN$ \\
\hline
\eqref{eq24} & $KN^{\mathsf d}(N^{\mathsf d}+1)$ \\
\hline
\eqref{eq25} & $NN^{\mathsf d}K+2N(N^{\mathsf d})^2+N^2+N$ \\
\hline
\textbf{Overall} &
$\begin{aligned}
&M\!(6K(N^{\mathsf d})^2\!\!+\!8N(N^{\mathsf d})^2
\!\!+\!\!3KNN^{\mathsf d}\!\!+\!\!7N^2\!\!+\!\!K^2\!\!+\!\!KN^{\mathsf d}
\!\!+\!2(N^{\mathsf d})^3 \\
&\quad \!\!\!\!+2KN\!\!+\!\!2K^2N^{\mathsf d}\!\!+\!\!6KN^2
\!\!+\!\!2K^2N\!\!+\!\!2K^2NN^{\mathsf d}+N)
\end{aligned}$ \\
\hline
\end{tabular}
\end{table}
Therefore, the computational complexity of the proposed deep-unfolded PGA method is $\mathcal{O}\big(MI(K^2NN^{\mathsf d}+KN^2)\big)$. In contrast, the PGA benchmark with line search has complexity $\mathcal{O}\big(MIL(K^2NN^{\mathsf d}+KN^2)\big)$, where $L$ denotes the average number of candidate step sizes evaluated during the line-search procedure. Moreover, an SCA-based benchmark requires solving a convex surrogate optimization problem at each iteration, leading to a complexity of $\mathcal{O}\big(I (MN^{\mathsf d}K+M(N^{\mathsf d})^2+2N)^3\big)$. Thus, PGA with line search mainly requires repeated gradient, projection, and objective evaluations, whereas SCA repeatedly constructs and solves convex surrogate subproblems. This highlights the advantage of the proposed deep-unfolded PGA method, which avoids both online line search and repeated convex subproblem solving. \CB

\section{Simulation Results}\label{simulations}
In this section, we present numerical results, averaged over 100 independent channel realizations, to validate the effectiveness of the proposed algorithm. We consider a BS equipped with a DMA-based architecture consisting of $N^{\mathsf{d}} = 5$ microstrips, each incorporating $N^{\mathsf{e}} = 10$ metamaterial elements. Furthermore, the central carrier frequency and system bandwidth are set to $f^{\mathsf c}=10$ GHz and $B=1$ GHz, respectively.  The convergence tolerance is set to $10^{-4}$. The BS is located at the origin and serves three users ($K=3$), each randomly distributed within the near-field region of the BS. The maximum transmit power at the BS and the number of subcarriers are set to $P^{\mathsf{max}}=10$ dB and $M=11$, respectively. Additionally,  the weighting factor is set to $\xi=0.4$. We conduct extensive simulations to evaluate the proposed deep-unfolded PGA framework against the following benchmarks:
\begin{itemize}
    \item \textbf{Classical PGA:} The digital beamformers, resonance frequencies, and damping factors are alternately updated using the derived gradients in \eqref{eq19}, \eqref{eq21}, and \eqref{eq17}, respectively, with fixed step sizes. 
    \item \textbf{PGA with line-search-based step-size selection}: The same alternating PGA framework is employed, but the step sizes are determined through a line-search procedure at each iteration rather than being fixed or learned. In this section, we refer to this method as PGA line-search.
    \item \textbf{MRT precoding:} At each iteration, the digital beamformers are constructed using MRT precoding, while the resonance frequencies and damping factors are iteratively updated according to the proposed deep-unfolded PGA method.
    \item \textbf{ZF precoding:} At each iteration, the digital beamformers are obtained by applying ZF precoding to the multi-user channel, while the DMA parameters are iteratively updated according to the proposed deep-unfolded PGA method.
\end{itemize}
Furthermore, to assess the impact of the frequency-selective (FS) Lorentzian model, its performance is compared with that of the frequency-flat (FF) Lorentzian model.
 
\begin{figure}
	\centering
	\includegraphics[scale=0.45]{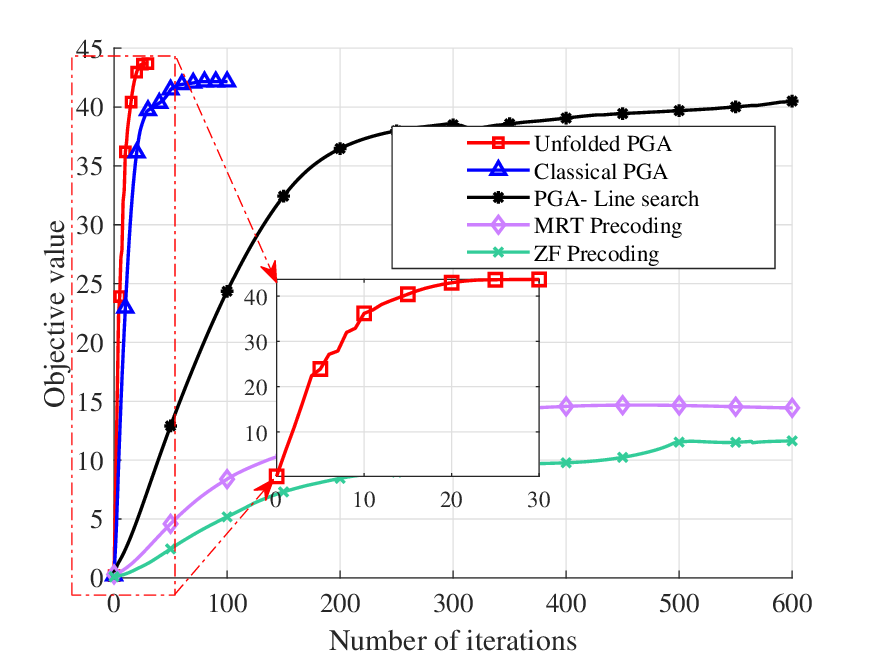}
	\caption{Convergence performance.}
    \label{Convergence}
\end{figure}
Fig. \ref{Convergence} compares the convergence behavior and achieved objective values of deep-unfolded PGA with four benchmark schemes under the frequency-selective Lorentzian model. The main observations are listed as follows:
\begin{itemize}
    \item \emph{Convergence speed:} Deep-unfolded PGA converges approximately $5$ and $20$ times faster than PGA line-search and classical PGA, respectively. It also achieves an approximately $20$-fold convergence-speed improvement over the MRT and ZF precodings.

    \item \emph{Achieved performance:} Deep-unfolded PGA improves the achieved objective value by approximately $3\%$ and $7\%$ over the PGA line-search and classical PGA, respectively. It also achieves approximately $3$-fold and $3.75$-fold higher objective values than the MRT and ZF precodings, respectively.
\end{itemize}
Furthermore, although the PGA line-search method achieves performance close to deep-unfolded PGA, it requires step-size optimization at every iteration. Specifically, it evaluates multiple candidate step sizes and repeatedly computes the objective value and projected solutions to select a suitable update. This additional step-size optimization increases the per-iteration computational complexity and latency. In contrast, the proposed deep-unfolded PGA method directly applies learned step sizes and therefore avoids per-iteration step-size search during inference.

\begin{figure}
    \centering
    \includegraphics[scale=0.45]{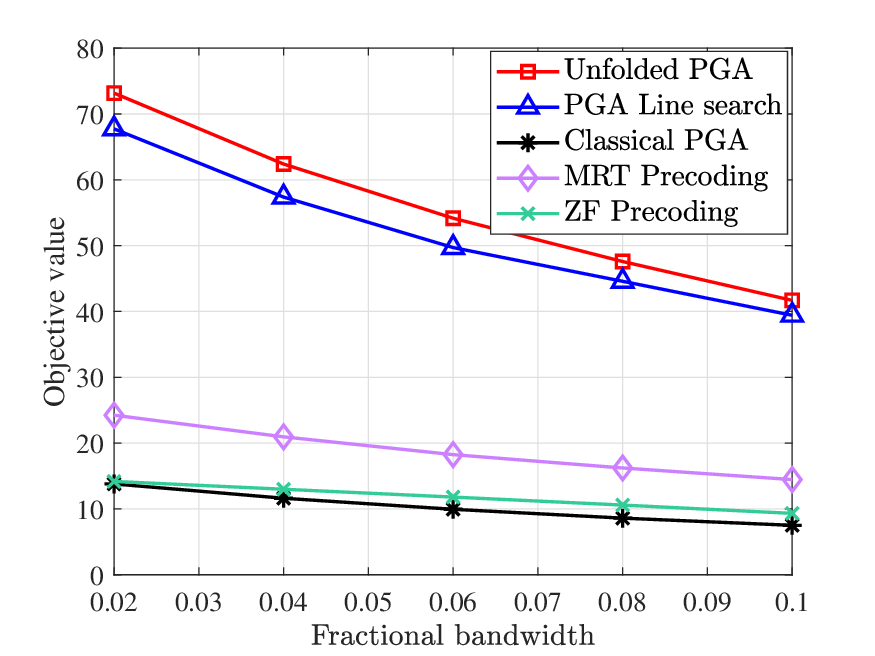}
\caption{Objective value versus fractional bandwidth under the frequency-selective Lorentzian model.}
\label{fractionalbandwidth1}
\end{figure}

\begin{figure}
    \centering
    \includegraphics[scale=0.45]{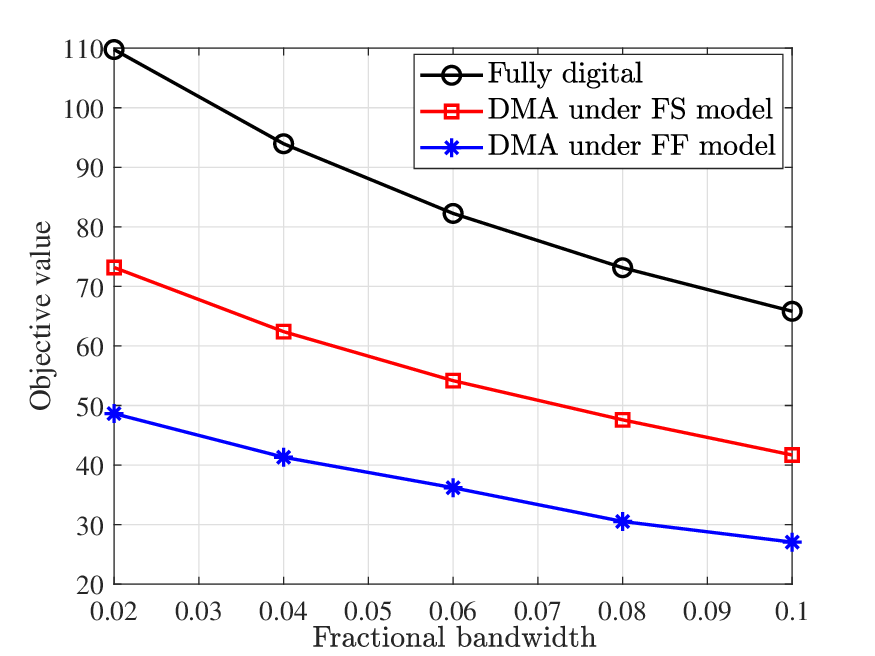}
    \caption{Objective value versus the fractional bandwidth.}
    \label{fractionalbandwidth2}
\end{figure}
Fig. \ref{fractionalbandwidth1} illustrates the impact of fractional bandwidth on the objective values achieved by the proposed deep-unfolded PGA method and the benchmark schemes under the frequency-selective Lorentzian model with the same number of iterations. As the fractional bandwidth increases, the objective value decreases for all schemes. The proposed deep-unfolded PGA method consistently achieves the highest objective value over the entire fractional-bandwidth range, followed closely by the PGA line-search method. At a fractional bandwidth of $0.02$, deep-unfolded PGA achieves approximately $7.4\%$ higher objective value than the PGA line-search method. Moreover, it achieves approximately $5.2$-fold, $3.0$-fold, and $5.2$-fold higher objective values than classical PGA, MRT precoding, and ZF precoding, respectively. Although the PGA line-search method achieves competitive performance, it incurs additional per-iteration complexity due to step-size search, whereas the proposed deep-unfolded PGA method directly applies learned step sizes and avoids this overhead.

Fig. \ref{fractionalbandwidth2} illustrates the impact of fractional bandwidth on the achieved objective value under two modeling frameworks for the frequency response of DMA elements, namely the frequency-selective Lorentzian model and its frequency-flat approximation. The main observations are as follows:
\begin{itemize}
    \item \emph{Modeling accuracy:} The results show that the frequency-flat approximation leads to noticeable performance degradation.  The performance gap between the
    two models can be primarily attributed to the limited modeling accuracy of the frequency-flat approximation. Specifically, the frequency-flat representation introduces phase deviations relative to the frequency-selective Lorentzian model, which becomes particularly problematic in wideband scenarios. In contrast, the nonlinear frequency-selective Lorentzian model explicitly incorporates both amplitude and phase variations into the resource allocation process according to the exact nonlinear physical model in \eqref{eq6}, thereby enabling more flexible and accurate optimization.

    \item \emph{Bandwidth sensitivity:} Quantitatively, over the considered fractional bandwidth range, the frequency-selective Lorentzian model exhibits only a two-fold reduction relative to its peak objective value, whereas the frequency-flat approximation experiences an approximately three-fold reduction. This further highlights the increasing inaccuracy of the frequency-flat approximation as the fractional bandwidth grows.
\end{itemize}

\begin{figure}
    \centering
    \includegraphics[scale=0.45]{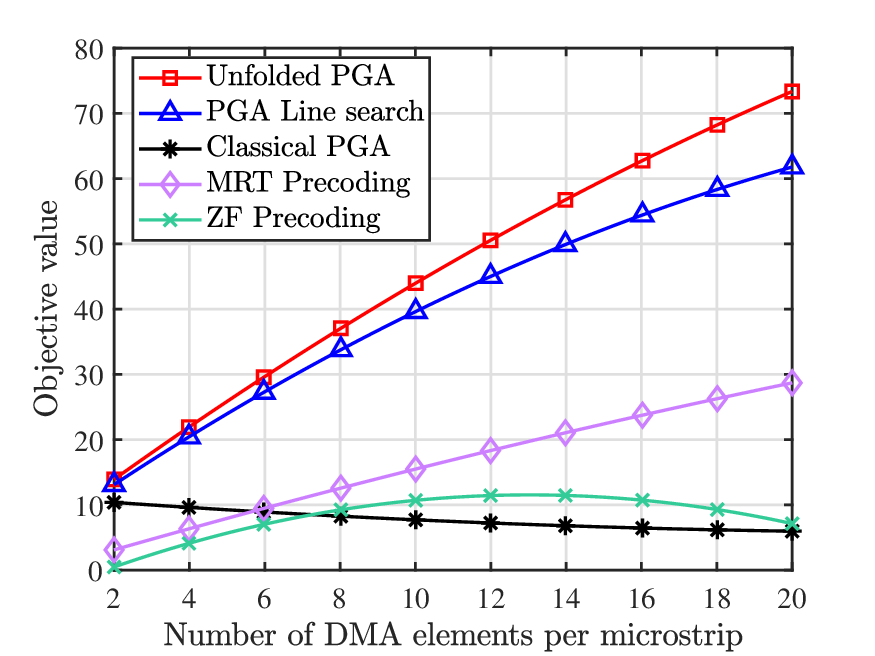}
    \caption{Objective value versus the number of elements per microstrip under the frequency-selective Lorentzian model.}
    \label{Antenn}
\end{figure}

Fig. \ref{Antenn} illustrates the impact of the number of DMA elements per microstrip on the objective values achieved by the proposed deep-unfolded PGA method and the benchmark schemes under the frequency-selective Lorentzian model with the same number of iterations. The proposed deep-unfolded PGA method consistently achieves the highest objective value across the entire range of DMA elements. As the number of DMA elements increases, the performance gap between the proposed deep-unfolded PGA method and the benchmark schemes becomes more pronounced, demonstrating its improved effectiveness in larger DMA configurations. For example, when $N^{\mathsf e}=10$, the proposed deep-unfolded PGA method achieves approximately $10\%$ higher objective value than the PGA line-search method. When $N^{\mathsf e}$ increases to $20$, this gain increases to approximately $18\%$.

\begin{figure}[b!]
    \centering
    \includegraphics[scale=0.45]{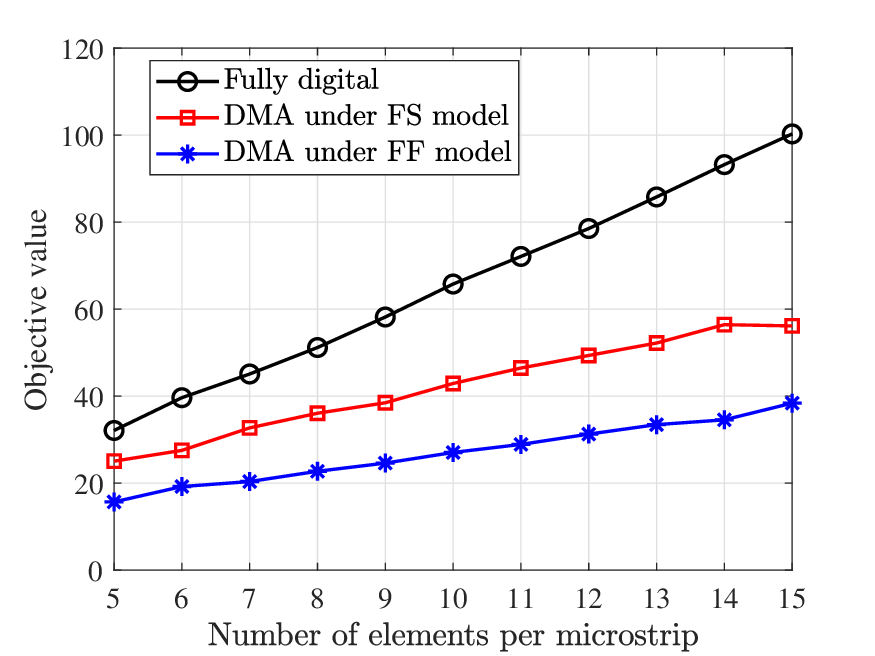}
    \caption{Objective value versus the number of DMA elements per microstrip.}
    \label{AntennFS}
\end{figure}
Fig. \ref{AntennFS} illustrates the impact of the number of elements per microstrip on the achieved objective value under the frequency-selective Lorentzian model and its frequency-flat approximation. Similar to Fig. \ref{fractionalbandwidth2}, the results show that the DMA-based architecture under the frequency-selective model consistently outperforms its frequency-flat approximation. This performance gap can be mainly attributed to the limited modeling accuracy of the frequency-flat approximation, which fails to capture the frequency-selective response of DMA elements and therefore leads to a noticeable degradation in the achieved objective value. Furthermore, as the number of elements increases, the performance gap between the frequency-selective model and its frequency-flat approximation becomes more pronounced. This demonstrates the importance of accurately modeling the frequency-selective response of DMA elements, especially in larger DMA configurations.
\begin{figure*}[t!]
    \vspace{-25pt}
    \centering

    \begin{subfigure}{0.32\textwidth}
        \centering
        \includegraphics[width=\linewidth]{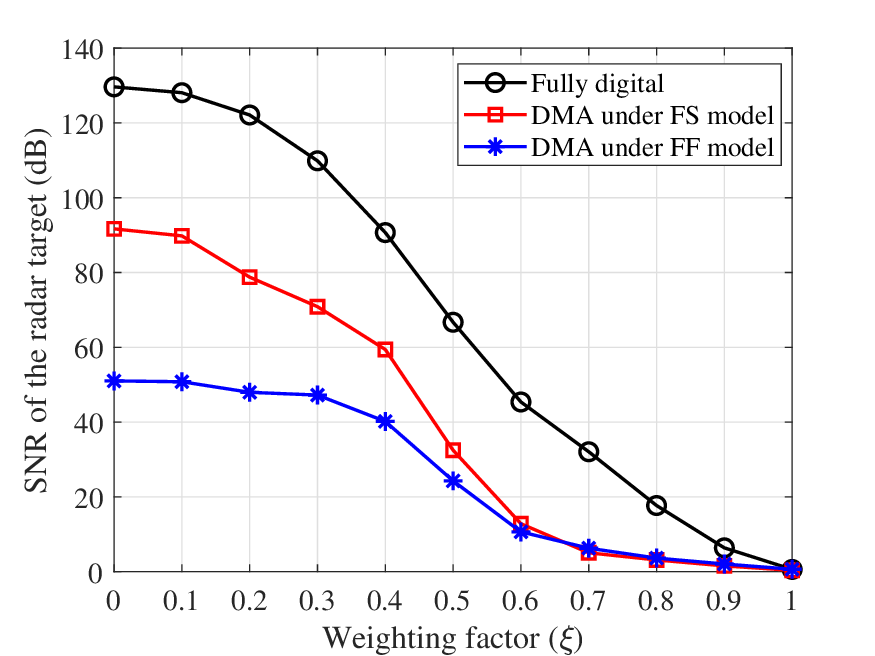}
        \caption{SNR of the radar target versus the weighting factor.}
         \label{radarxi}
    \end{subfigure}
    \hfill
    \begin{subfigure}{0.32\textwidth}
        \centering
        \includegraphics[width=\linewidth]{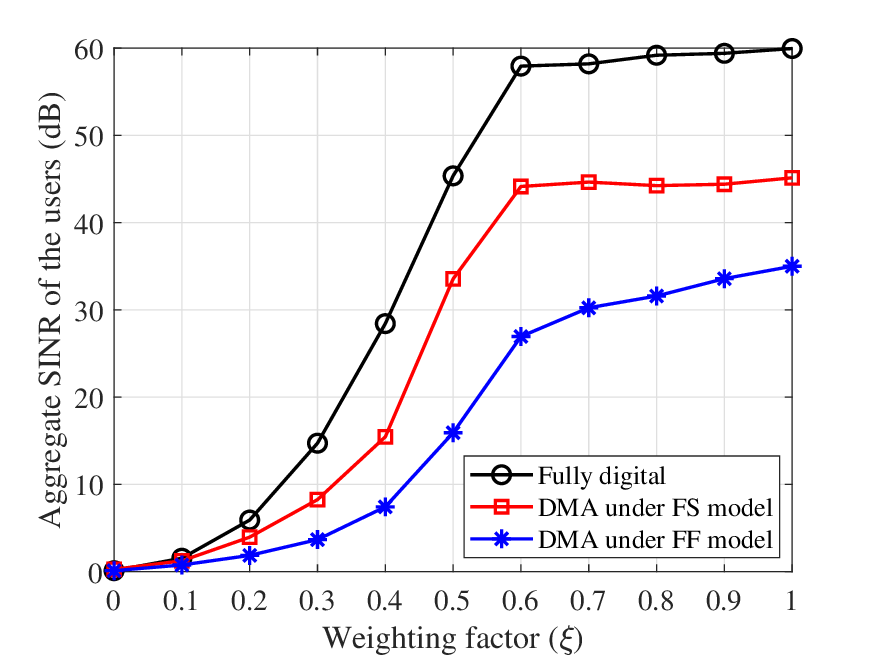}
        \caption{Aggregate SINR of the users versus the weighting factor.}
        \label{userxi}
    \end{subfigure}
    \hfill
    \begin{subfigure}{0.32\textwidth}
        \centering
        \includegraphics[width=\linewidth]{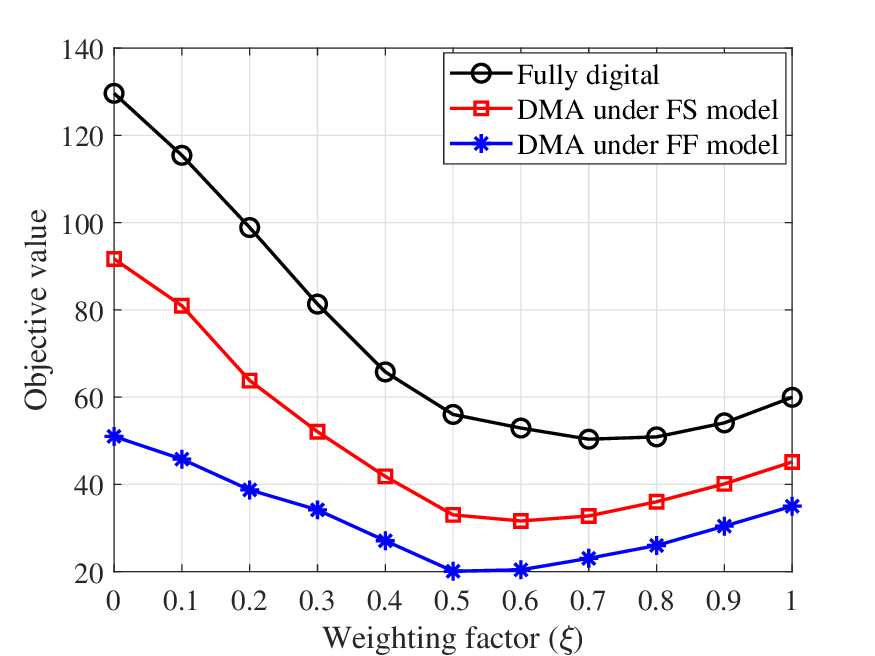}
        \caption{Objective value versus the weighting factor.}
        \label{objectivexi}
    \end{subfigure}
    \caption{System performance versus the weighting factor ($\xi$).}
    \label{fig_weighting}
\end{figure*}

Figure \ref{fig_weighting} illustrates the system performance as a function of the weighting factor $\xi$  for the proposed deep-unfolded PGA model under the fully digital architecture and the DMA-based architectures, considering both the frequency-selective Lorentzian model and its frequency-flat approximation. The main observations are as follows:
\begin{itemize}
    \item Fig. \ref{radarxi} shows the SNR of the radar target versus the weighting factor $\xi$. As  $\xi$ increases, the radar target SNR decreases for all architectures, indicating that larger values of $\xi$ reduce the priority given to radar sensing performance. The fully digital architecture achieves the highest radar SNR, followed by the DMA under the frequency-selective Lorentzian model, while the DMA under the frequency-flat approximation gives the lowest radar SNR.

    \item Fig. \ref{userxi} presents the aggregate SINR of the users versus the weighting factor $\xi$. In contrast to the radar SNR, the aggregate user SINR increases as $\xi$ increases. This indicates that larger values of $\xi$ give more priority to communication performance. Again, the fully digital architecture provides the best performance, and the DMA under the frequency-selective Lorentzian model outperforms the frequency-flat approximation.

    \item Fig. \ref{objectivexi} shows the objective value versus the weighting factor $\xi$. The objective value first decreases and then increases as $\xi$ grows, reflecting the balance between the radar sensing and communication terms in the objective function. The fully digital architecture has the highest objective value, while the frequency-selective DMA model achieves better objective values than the frequency-flat approximation.
\end{itemize}
Overall, these results demonstrate the sensing--communication tradeoff controlled by $\xi$ and show that the frequency-flat approximation model leads to noticeable performance degradation in the DMA-based wideband beam management problem.

\begin{figure}
    \centering
    \includegraphics[scale=0.45]{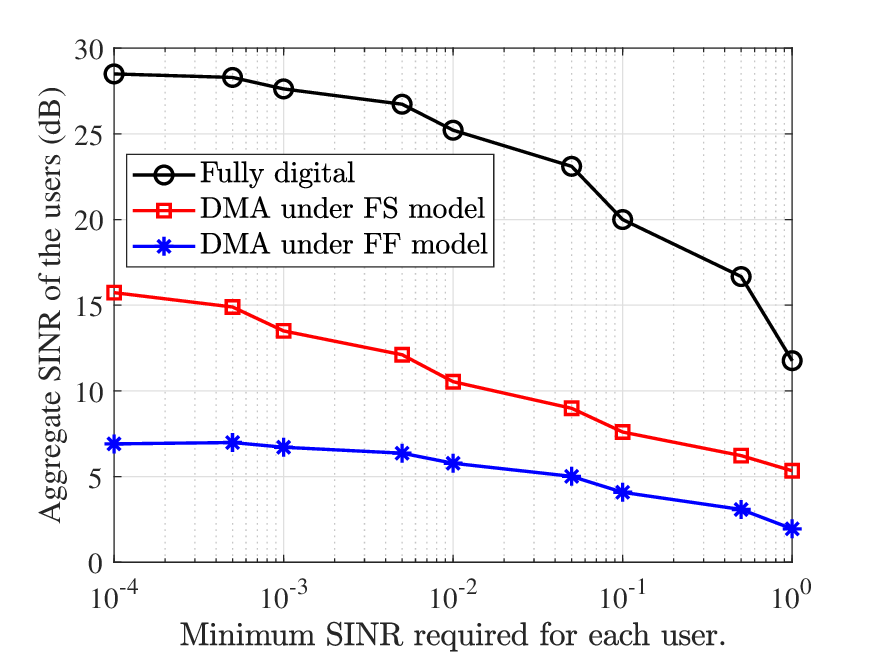}
    \caption{Aggregate SINR of the users versus the minimum SINR required for each user.}
    \label{minsinr}
\end{figure}


Fig. \ref{minsinr} shows the aggregate users’ SINR performance versus the minimum SINR requirement for each user. Increasing the minimum SINR requirement makes the system design more restrictive, thereby reducing the achievable users’ SINR for all schemes. The frequency-selective model outperforms the frequency-flat model over the entire range. For example, the frequency-selective model decreases from approximately $16$ dB to about $5.5$ dB, while the frequency-flat model decreases from around $7$ dB to nearly $2$ dB.

\section{Conclusion}\label{conclusion}
In this paper, we proposed a beamforming design for DMA-based wideband ISAC systems under a frequency-selective Lorentzian response model, which captures the frequency-dependent magnitude and phase responses of DMA elements in wideband operation. We formulated a communication-and-sensing optimization problem to balance the aggregate users’ SINR and the radar-target SNR by jointly optimizing the digital beamforming matrices, resonance frequencies, and damping factors. To solve the resulting non-convex optimization problem, we first developed an alternating PGA-based framework with closed-form gradients. We then constructed a deep-unfolded PGA framework that learns and tunes the step-size parameters from data. Simulation results showed that the frequency-selective Lorentzian model achieves approximately a $20\%$ performance improvement over the frequency-flat approximation in our scenario. The deep-unfolded PGA method also converges significantly faster than PGA with line-search-based step-size selection and classical PGA. Furthermore, deep-unfolded PGA achieves up to a 7\% higher objective value than the PGA-based benchmarks. These results demonstrated that the frequency-selective Lorentzian model, combined with the deep-unfolded PGA-based framework, can substantially improve wideband ISAC beamforming while reducing computational overhead.

\appendices
\section{}\label{AppendixA}
By setting $\mathbf{q}_m = \text{vec}(\mathbf{Q}_m)\in \mathbb{C}^{N N^{\mathsf d}\times1}$, the SINR of the $k$-th communication user at the $m$-th subcarrier is reformulated as follows:
\begin{equation}\label{eq27}
    \gamma^{\mathsf u}_{m,k} = \frac{\vert \mathbf{z}_{m,k,k}^{\mathsf u}\mathbf{q}_m \vert^2}{\sum_{i=1,i\neq k}^{K} \vert \mathbf{z}_{m,i,k}^{\mathsf u}\mathbf{q}_m \vert^2 + \sum_{j=1}^{N^{\mathsf d}} \vert \mathbf{z}_{m,j,k}^{\mathsf r}\mathbf{q}_m \vert^2+\sigma_{m,k}^2},
\end{equation}
where $\mathbf{z}^{\mathsf r}_{m,j,k} = (\mathbf{w}^{\mathsf r}_{m,j})^T \otimes (\mathbf{g}^{\mathsf u}_{m,k})^H\mathbf{H}_m\in \mathbb{C}^{1 \times N N^{\mathsf d}}$. Furthermore, $\mathbf{z}_{m,i,k}^{\mathsf u}$ is defined in Theorem 1. In addition, \eqref{eq27} is reformulated as follows:
\begin{equation}\label{eq27.1}
    \gamma^{\mathsf u}_{m,k} = \frac{\vert \hat{\mathbf{z}}_{m,k,k}^{\mathsf u}\hat{\mathbf{q}}_m \vert^2}{\sum_{i=1,i\neq k}^{K} \vert \hat{\mathbf{z}}_{m,i,k}^{\mathsf u}\hat{\mathbf{q}}_m \vert^2 + \sum_{j=1}^{N^{\mathsf d}} \vert \hat{\mathbf{z}}_{m,j,k}^{\mathsf r}\hat{\mathbf{q}}_m \vert^2+\sigma_{m,k}^2}.
\end{equation}
where $\hat{\mathbf{z}}_{m,j,k}^{\mathsf r}$ denotes the modified version of ${\mathbf{z}}_{m,j,k}^{\mathsf r}$. Furthermore, $\hat{\mathbf{q}}_m$ and $\hat{\mathbf{z}}_{m,j,k}^{\mathsf u}$ are defined in Theorem 1. By imposing $\text{tr}(\mathbf{Q}_m^H \mathbf{B}_m \mathbf{E}_m \mathbf{Q}_m) = \hat{\mathbf{q}}^H_m \mathbf{C}_m \hat{\mathbf{q}}_m$, in which $\mathbf{C}_m = (\mathbf{B}_m\otimes \mathbf{I}_{N^{\mathsf e}})\odot \mathbf{E}_m$, the SNR of the radar target at the $m$-th subcarrier is reformulated as follows:
\begin{equation}\label{eq28}
    \gamma^{\mathsf r}_m= \frac{1}{N\tilde{\sigma}_m^2} \hat{\mathbf{q}}^H_m \left[(\mathbf{B}_m\otimes \mathbf{I}_{N^{\mathsf e}})\odot\mathbf{E}_m\right]\hat{\mathbf{q}}_m,
\end{equation}
where $\mathbf{B}_m$ and $\mathbf{E}_m$ are defined in Theorem 1. 
The gradients of $\gamma^{\mathsf u, \mathsf {tot}}$  with respect to ${\mathbf f}^{\mathsf R}$ are obtained as follows \cite{derive}:
\begin{equation}\label{eqA1}
\begin{aligned}
\nabla_{\mathbf{f}^{\mathsf R}}\gamma^{\mathsf u,\mathsf{tot}}
= \Re\Big\lbrace\sum_{m=1}^{M}\sum_{k=1}^{K}\frac{\partial\hat{\mathbf q}_m}
     {\partial\mathbf f^{\mathsf R}}\big(\frac{\partial\gamma^{\mathsf u}_{m,k}}
     {\partial\hat{\mathbf q}^*_m}\big)^*
\Big\rbrace,
\end{aligned}
\end{equation}
where $\frac{\partial\hat{\mathbf q}_m}{\partial\mathbf f^{\mathsf R}} = \text{diag} \left[\frac{\partial\hat{q}_{m,1,1}}{\partial f^{\mathsf R}_{1,1}}, \frac{\partial\hat{q}_{m,1,2}}{\partial f^{\mathsf R}_{1,2}}, \dots, \frac{\partial\hat{q}_{m,N^{\mathsf d}, N^{\mathsf e}}}{\partial f^{\mathsf R}_{N^{\mathsf d}, N^{\mathsf e}}} \right]$. Using \eqref{eq6}, we can obtain $\frac{\partial\hat{ q}_{m,i,l}}{\partial f^{\mathsf R}_{i,l}} = -\frac{2f^{\mathsf R}_{i,l}(F_{i,l} f_m^2)}{\left((f_{i,l}^{\mathsf R})^2 - f_m^2 - j\kappa_{i,l} f_m\right)^2}$. In addition, using \eqref{eq27.1}, we obtain $\frac{\partial\gamma^{\mathsf u}_{m,k}}
{\partial\hat{\mathbf q}^*_m} = \frac{\hat{\mathbf{u}}_{m,k} A_{m,k} - D_{m,k} \hat{\mathbf{t}}_{m,k}}{A_{m,k}^2}$, where $\hat{\mathbf{u}}_m$, $A_{m,k}$, and $D_{m,k}$ are defined in Theorem 1. Since $\frac{\partial\hat{\mathbf q}_m}{\partial\mathbf f^{\mathsf R}}$ and $\frac{\partial\gamma^{\mathsf u}_{m,k}}
{\partial\hat{\mathbf q}_m}$ are associated element-wise with the same DMA elements, their product is expressed using the Hadamard product operator in Theorem 1. Therefore, $\nabla_{\mathbf{f}^{\mathsf R}}\gamma^{\mathsf u,\mathsf{tot}}$ is obtained as in \eqref{eq29}. 
The gradients of $\gamma^{\mathsf r, \mathsf {tot}}$  with respect to ${\mathbf f}^{\mathsf R}$ are obtained as follows \cite{derive}:
\begin{equation}\label{eqA2}
\begin{aligned}
\nabla_{\mathbf{f}^{\mathsf R}}\gamma^{\mathsf r,\mathsf{tot}}
&=\Re\Big\lbrace \sum_{m=1}^{M}
\frac{\partial\hat{\mathbf q}_m}
     {\partial\mathbf f^{\mathsf R}}\left(\frac{\partial\gamma^{\mathsf r}_m}
     {\partial\hat{\mathbf q}_m^*}\right)^*\Big\rbrace.                              
\end{aligned}
\end{equation}

For a quadratic function $f(\mathbf{q}_m)=\mathbf{q}_m^H\mathbf{F}_m\mathbf{q}_m$, the Wirtinger derivative is $\frac{\partial f}{\partial \mathbf{q}^*_m} = \mathbf{F}_m\mathbf{q}_m$. Therefore, by regarding all terms between $\mathbf{q}_m^H$ and $\mathbf{q}_m$ in \eqref{eq28} as the constant matrix $\mathbf{F}_m$, $ \frac{\partial\gamma^{\mathsf r,\mathsf{tot}}}
     {\partial\hat{\mathbf q}_m^*}$ is obtained as follows \cite{derive}:
\begin{equation}\label{XXXX}
    \frac{\partial\gamma^{\mathsf r}_m}
     {\partial\hat{\mathbf q}_m^*} = \frac{2}{N\tilde{\sigma}_m^2}\left[(\mathbf{B}_m\otimes \mathbf{I}_{N^{\mathsf e}})\odot\mathbf{E}_m\right]\hat{\mathbf{q}}_m.
\end{equation}

By substituting \eqref{XXXX} and $\frac{\partial\hat{\mathbf q}_m}{\partial\mathbf f^{\mathsf R}}$ from \eqref{eqA1}, into \eqref{eqA2}, $\nabla_{\mathbf{f}^{\mathsf R}}\gamma^{\mathsf r,\mathsf{tot}}$ is obtained as in \eqref{eq30}. Finally, $\nabla_{\mathbf f^{\mathsf R}}\Lambda$ is obtained as follows:
\begin{equation}\label{eqA3}
\begin{aligned}
\nabla_{\mathbf f^{\mathsf R}}\Lambda
&=\Re \Big\lbrace
\sum_{m=1}^{M}\sum_{k=1}^{K}
\frac{1
}{
\sum_{\ell=1}^{M}\gamma_{\ell,k}^{\mathsf u}
-\Gamma_k^{\mathsf{min}}
}\left(\frac{\partial\hat{\mathbf q}_m} {\partial\mathbf f^{\mathsf R}}\left(\frac{\partial\gamma^{\mathsf u}_m} {\partial\hat{\mathbf q}_m^*}\right)^*\right)\Big\rbrace                              \\
&\!\!\!\!=
2\sum_{k=1}^{K}
\frac{
\Re\Bigg\{
\displaystyle\sum_{m=1}^{M}
\mathbf b_m\odot
\left(\frac{
\hat{\mathbf u}_{m,k}A_{m,k}
-D_{m,k}\hat{\mathbf t}_{m,k}
}{
A_{m,k}^{2}
}\right)^*
\Bigg\}
}{
\displaystyle\sum_{\ell=1}^{M}\gamma_{\ell,k}^{\mathsf u}
-\Gamma_k^{\mathsf{min}}
}.
\end{aligned}
\end{equation}
where $\mathbf{b}_m$ is defined in Theorem 1. Furthermore, $\Gamma_k^{\mathsf{min}}$ is defined in \eqref{eq15}.
The proof is thus completed.

\section{}\label{AppendixB}
The gradients of $\gamma^{\mathsf u, \mathsf {tot}}$  with respect to $\boldsymbol{\kappa}$ is obtained as follows \cite{derive}:

\begin{equation}\label{eqB1}
\begin{aligned}
\nabla_{\boldsymbol{\kappa}}\gamma^{\mathsf u,\mathsf{tot}}
= \Re\Big\lbrace\sum_{m=1}^{M}\sum_{k=1}^{K}\frac{\partial\hat{\mathbf q}_m}
     {\partial \boldsymbol{\kappa}}\big(\frac{\partial\gamma^{\mathsf u}_{m,k}}
     {\partial\hat{\mathbf q}^*_m}\big)^*
\Big\rbrace.
\end{aligned}
\end{equation}
Similar to Appendix  \ref{AppendixA}, we have $\frac{\partial\hat{\mathbf q}_m}{\partial \boldsymbol{\kappa}} = \text{diag} \left[\frac{\partial\hat{q}_{m,1,1}}{\partial \kappa_{1,1}}, \frac{\partial\hat{q}_{m,1,2}}{\partial \kappa_{1,2}}, \dots, \frac{\partial\hat{q}_{m,N^{\mathsf d}, N^{\mathsf e}}}{\partial \kappa_{N^{\mathsf d}, N^{\mathsf e}}} \right]$. Using \eqref{eq6}, we can obtain $\frac{\partial\hat{ q}_{m,i,l}}{\partial \kappa_{i,l}} = \frac{jF_{i,l} f_m^3}{\left((f_{i,l}^{\mathsf R})^2 - f_m^2 - j\kappa_{i,l} f_m\right)^2}$. By substituting $\frac{\partial\hat{\mathbf q}_m}{\partial \boldsymbol{\kappa}}$ and $\frac{\partial\gamma^{\mathsf u,\mathsf{tot}}}
{\partial\hat{\mathbf q}_m^*}$, obtained from Appendix \ref{AppendixA}, into \eqref{eqB1}, $\nabla_{\boldsymbol{\kappa}}\gamma^{\mathsf u,\mathsf{tot}}$ is obtained as in \eqref{eq32}. Furthermore, the gradients of $\gamma^{\mathsf r, \mathsf {tot}}$  with respect to $\boldsymbol{\kappa}$ is obtained as follows \cite{derive}:

\begin{equation}\label{eqB2}
\begin{aligned}
\nabla_{\boldsymbol{\kappa}}\gamma^{\mathsf r,\mathsf{tot}}
&= \Re\Big\lbrace \sum_{m=1}^{M}
\frac{\partial\hat{\mathbf q}_m}
     {\partial \boldsymbol{\kappa}}\left(\frac{\partial\gamma^{\mathsf r}_m}
     {\partial\hat{\mathbf q}_m^*}\right)^*\Big\rbrace.                                  
\end{aligned}
\end{equation}

Using \eqref{XXXX} and $\frac{\partial\hat{\mathbf q}_m}{\partial\boldsymbol{\kappa}}$ from \eqref{eqB1}, $\nabla_{\boldsymbol{\kappa}}\gamma^{\mathsf r,\mathsf{tot}}$ is obtained as in \eqref{eq33}. Finally, $\nabla_{\boldsymbol{\kappa}}\Lambda$ is obtained as follows: 

\begin{equation}\label{eqB3}
\begin{aligned}
\nabla_{\boldsymbol{\kappa}}\Lambda
&=
\sum_{m=1}^{M}\sum_{k=1}^{K}
\frac{1
}{
\sum_{\ell=1}^{M}\gamma_{\ell,k}^{\mathsf u}
-\Gamma_k^{\mathsf{min}}
}\left(\frac{\partial\hat{\mathbf q}_m} {\partial\boldsymbol{\kappa}} \left(\frac{\partial\gamma^{\mathsf u}_{m,k}} {\partial\hat{\mathbf q}_m^*}\right)^* \right)                              \\
&\!\!\!=
2\sum_{k=1}^{K}
\frac{
\Re\Bigg\{
\displaystyle\sum_{m=1}^{M}
\mathbf p_m\odot
\left(\frac{
\hat{\mathbf u}_{m,k}A_{m,k}
-D_{m,k}\hat{\mathbf t}_{m,k}
}{
A_{m,k}^{2}
}\right)^*
\Bigg\}
}{
\displaystyle\sum_{\ell=1}^{M}\gamma_{\ell,k}^{\mathsf u}
-\Gamma_k^{\mathsf{min}}
}.
\end{aligned}
\end{equation}
where $\mathbf{p}_m$ is defined in Theorem 2.
The proof is thus completed.

\section{}\label{AppendixC}
To derive the gradient of $\gamma^{\mathsf u, \mathsf{tot}}$ with respect to $\mathbf{w}_{m,k}^{\mathsf u}$, we separately consider the contribution of the desired SINR term of user $k$ and the interference terms of the other users. For the desired SINR term of user $k$, the gradient with respect to its own communication beamformer is obtained as follows \cite{derive}:
\begin{equation}\label{appC1}
    \nabla_{\mathbf{w}_{m,k}^{\mathsf u}}
\gamma^{\mathsf u}_{m,k}=
\frac{
    2\mathbf{v}_{m,k}^{H}\mathbf{v}_{m,k}
    \mathbf{w}_{m,k}^{\mathsf u}
}{
    A_{m,k}
}.
\end{equation}

For the SINR term of user $j$, $j\neq k$, the beamformer of user $k$ appears only in the interference term. Therefore, its contribution to the gradient is obtained separately as follows:
\begin{equation}\label{appC2}
    \nabla_{\mathbf{w}_{m,k}^{\mathsf u}}
\gamma^{\mathsf u}_{m,j}
=
-2
\frac{
    \left|
    \mathbf{v}_{m,j}\mathbf{w}_{m,j}^{\mathsf u}
    \right|^{2}
}{
    A_{m,j}^{2}
}
\mathbf{v}_{m,j}^{H}\mathbf{v}_{m,j}
\mathbf{w}_{m,k}^{\mathsf u}, \quad j\neq k,
\end{equation}

By combining the desired-term contribution in \eqref{appC1} with the interference contributions from all other users in \eqref{appC2}, the desired gradient expression in \eqref{eq23} is obtained. Furthermore, \eqref{eq24} is obtained using an approach analogous to that employed in deriving \eqref{appC2}. Finally, following a procedure similar to that used for \eqref{XXXX},  $\nabla_{\mathbf{W}_m}
\gamma^{\mathsf r,\mathsf{tot}}$ is obtained as follows:
\begin{equation}
    \nabla_{\mathbf{W}_m}
\gamma^{\mathsf r,\mathsf{tot}}
=
\frac{2}{N\tilde{\sigma}_m^2}
\mathbf{Q}_m^{H}\mathbf{H}_m^{H}
(\mathbf{G}_m^{\mathsf r})^{H}\mathbf{G}_m^{\mathsf r}
\mathbf{H}_m\mathbf{Q}_m\mathbf{W}_m.
\end{equation}
The proof is thus completed.

\bibliographystyle{IEEEtran}
\bibliography{bib.bib}
\end{document}